\documentclass[fleqn,usenatbib]{mnras}

\usepackage{newtxtext,newtxmath}
\usepackage[T1]{fontenc}

\DeclareRobustCommand{\VAN}[3]{#2}
\let\VANthebibliography\thebibliography
\def\thebibliography{\DeclareRobustCommand{\VAN}[3]{##3}\VANthebibliography}

\usepackage{graphicx}	% Including figure files
\usepackage{amsmath}	% Advanced maths commands

\usepackage[utf8]{inputenc}

\usepackage{makecell}

\title[Decoupling outflow and disk kinematics]{Decoupling the AGN outflow and star-forming disk kinematics in the nuclear region of NGC~7582 with JWST NIRSpec and MIRI/MRS}

\author[O. Veenema et al.]{
Oscar Veenema,$^{1}$\thanks{E-mail: oscar.veenema@physics.ox.ac.uk}
Niranjan Thatte,$^{1}$
Dimitra Rigopoulou,$^{1, 2}$
Ismael Garc\'{i}a-Bernete,$^{3}$
\newauthor
Almudena Alonso-Herrero,$^{3}$
Miguel Pereira-Santaella,$^{4}$
Anelise Audibert,$^{5, 6}$
Enrica Bellocchi,$^{7, 8}$
\newauthor
Andrew J. Bunker,$^{1}$
Steph Campbell,$^{9}$
Francoise Combes,$^{10}$
Ric I. Davies,$^{11}$
Fergus R. Donnan,$^{12}$
\newauthor
Santiago Garc\'{i}a-Burillo,$^{13}$
Omaira Gonzalez Martin,$^{14}$
Laura Hermosa Mu{\~n}oz,$^{3}$
Erin K. S. Hicks,$^{15, 16, 17}$
\newauthor
Sebastian F. Hoenig,$^{18}$
Alvaro Labiano,$^{19}$
Nancy A. Levenson,$^{20}$
Chris Packham,$^{17, 21}$
\newauthor
Cristina Ramos Almeida,$^{5, 6}$
Claudio Ricci,$^{22, 23}$
Rogemar A. Riffel,$^{24, 25}$
David Rosario,$^{9}$
\newauthor
Taro Shimizu,$^{11}$
Lulu Zhang$^{17}$
\\
$^{1}$Department of Physics, University of Oxford, Keble Road, Oxford, OX1 3RH, UK\\
$^{2}$School of Sciences, European University Cyprus, Diogenes street, Engomi, 1516 Nicosia, Cyprus\\
$^{3}$Centro de Astrobiolog\'{i}a (CAB), CSIC-INTA, Camino Bajo del 497 Castillo s/n, E-28692 Villanueva de la Ca{\~n}ada, Madrid, Spain\\
$^{4}$Instituto de F\'{i}sica Fundamental, CSIC, Calle Serrano 123, 28006 Madrid, Spain\\
$^{5}$Instituto de Astrof\'{i}sica de Canarias, Calle V\'{i}a L\'{a}ctea, s/n, E-38205, La Laguna, Tenerife, Spain\\
$^{6}$Departamento de Astrof\'{i}sica, Universidad de La Laguna, E-28206, La Laguna, Tenerife, Spain\\
$^{7}$Departmento de F\'{i}sica de la Tierra y Astrof\'{i}sica, Fac. de CC F\'{i}sicas, Universidad Complutense de Madrid, E-28040 Madrid, Spain\\
$^{8}$Instituto de F\'isica de Part\'iculas y del Cosmos IPARCOS, Fac. CC. F\'isicas, Universidad Complutense de Madrid, 28040 Madrid, Spain\\
$^{9}$School of Mathematics, Statistics and Physics, Newcastle University, Newcastle upon Tyne NE1 7RU, UK\\
$^{10}$Observatoire de Paris, LUX, Coll\`ege de France, CNRS, PSL University, Sorbonne University, 75014, Paris, France\\
$^{11}$Max Planck Institute for extraterrestrial Physics, Giessenbachstrasse 1, 85748, Garching, Germany\\
$^{12}$Department of Astronomy \& Astrophysics, University of California, San Diego, La Jolla,
CA 92093, USA\\
$^{13}$Observatorio Astron\'{o}mico Nacional (OAN-IGN)-Observatorio de Madrid, Alfonso XII, 3, 28014 Madrid, Spain\\
$^{14}$Instituto de Radioastronom\'ia y Astrof\'isica (IRyA), Universidad Nacional Autonoma de Mexico, Mexico\\
$^{15}$Department of Physics \& Astronomy, University of Alaska Anchorage, Anchorage, AK 99508-4664, USA\\
$^{16}$Department of Physics, University of Alaska Fairbanks, Fairbanks, AK 99775-5920, USA\\
$^{17}$Department of Physics and Astronomy, The University of Texas at San Antonio, 1 UTSA Circle, San Antonio, TX 78249, USA\\
$^{18}$School of Physics and Astronomy, University of Southampton, Southampton, SO17 1BJ, UK\\
$^{19}$Telespazio UK for the European Space Agency (ESA), ESAC, Camino Bajo del Castillo s/n, 28692 Villanueva de la Ca\~nada, Madrid, Spain\\
$^{20}$Space Telescope Science Institute, San Martin Drive, Baltimore, MD 21218, USA\\
$^{21}$National Astronomical Observatory of Japan, National Institutes of Natural Sciences (NINS), 2-21-1 Osawa, Mitaka, Tokyo 181-8588, Japan\\
$^{22}$Department of Astronomy, University of Geneva, ch. d'Ecogia 16, 1290, Versoix, Switzerland\\
$^{23}$Instituto de Estudios Astrof\'isicos, Facultad de Ingenier\'ia y Ciencias, Universidad Diego Portales, Av. Ej\'ercito Libertador 441, Santiago, Chile\\
$^{24}$Departamento de F\'isica, CCNE, Universidade Federal de Santa Maria, 97105-900 Santa Maria, RS, Brazil\\
$^{25}$Centro de Astrobiolog\'ia (CAB), CSIC-INTA, Ctra. de Ajalvir km 4, Torrej\'on de Ardoz, E-28850, Madrid, Spain\\
}

\date{Accepted XXX. Received YYY; in original form ZZZ}

\pubyear{\the\year{}}

\begin{document}
\label{firstpage}
\pagerange{\pageref{firstpage}--\pageref{lastpage}}
\maketitle

\begin{abstract}

We present a detailed study of the inner regions of NGC~7582, a nearby Seyfert~2 galaxy, from the Galaxy Activity, Torus and Outflow Survey (GATOS). The galaxy hosts a circumnuclear star-forming disk and an AGN-driven biconical ionised outflow. Using JWST NIRSpec and MIRI/MRS integral-field spectroscopy, we analyse ionic emission lines spanning a wide range of ionisation potentials (IPs, $\sim 8$–$126$ eV). Gaussian line-profile fitting reveals kinematic stratification: low-IP species ($\lesssim 20$ eV; e.g., [Fe II], [Ar II], [Ne II]) trace ordered disk rotation with PA $\sim -12 \pm 3^\circ$, while high-IP species ($\gtrsim 35$ eV; e.g., [O IV], [Mg IV], [Ne V]) follow the outflow with PA $\sim 54 \pm 10^\circ$. Outflowing gas exhibits systematically higher velocity dispersions ($119 \pm 13$ km/s) than the disk ($78 \pm 11$ km/s), consistent with turbulent or bulk motions. Intermediate-IP lines, [S III], [Ar III], and [Ne III], show contributions from both components, with the outflow characterised by higher dispersion, lower amplitude, and higher velocities in double-Gaussian fits. For these lines, a thin inclined disk plus one-dimensional outflow model enables robust separation and quantification of the disk and outflow velocity fields. The outflow is consistent with a hollow bicone capable of accelerating gas beyond the local escape velocity, implying most material is unlikely to be re-accreted. The ionisation cone opening angle shows no dependence on IP, indicating the AGN torus polar regions are largely unobscured. Our study provides new insights into AGN-driven outflows and circumnuclear disk dynamics, offering a framework to disentangle overlapping ISM kinematics in nearby active galaxies.

\end{abstract}


%%%%%%%%%%%%%%%%%%%%%%%%%%%%%%%%%%%%%%%%%%%%%%%%%%

%%%%%%%%%%%%%%%%% BODY OF PAPER %%%%%%%%%%%%%%%%%%

\section{Introduction}

In the central regions of galaxies, the kinematics of the gas in the interstellar medium (ISM) can deviate markedly from the ordered disk-like rotation observed on galactic scales (e.g., \citealt{mingozzi2019magnum, garcia2021multiphase, heckler2022ifu, esparza2025molecular, lin2025gas}). These inner zones are dynamically complex, shaped by processes such as active galactic nuclei (AGN) feedback, winds from nuclear and circumnuclear star-formation, and gravitational interactions from mergers, to list but a few. AGN can launch powerful winds and jets that ionise and heat the surrounding ISM, while nuclear or circumnuclear starbursts generate supernova-driven feedback. Together these mechanisms can produce gaseous streaming motions (inflows and outflows) (e.g., \citealt{cicone2014massive, harrison2014kiloparsec}), shocks (e.g., \citealt{huang2022chemical, davies2024gatos}), and enhanced turbulence (e.g., \citealt{mullaney2013narrow, venturi2021magnum, ulivi2024feedback}) within the ISM. This often creates a complex kinematic environment which can directly affect the wider structure and composition of the galaxy as a whole \citep{guillard2015exceptional, harrison2024observational}, highlighting the interplay between AGN feedback and galaxy evolution.

ISM kinematics can be studied through the analysis of emission lines, each tracing different phases, stratified by quantities such as gas temperature and level of ionisation. Molecular gas, primarily composed of molecular hydrogen (H$_2$) and carbon monoxide (CO), is particularly well-studied. CO traces the coldest molecular phase \citep{aravena2010cold, almeida2022diverse, jones2023evidence}, while H$_2$ spans a broader and warmer temperature range \citep{rigopoulou2002iso, roussel2007warm, 2016ApJ...830...18T}. Owing to its abundance, H$_2$ is especially valuable for probing molecular gas on nuclear and circumnuclear scales \citep{rosenberg2013excitation, pereira2022low, costa2024blowing, davies2024gatos, riffel2025blowing, esparza2025molecular}. Beyond the molecular phase, atomic hydrogen emits a series of recombination lines which allows us to trace star-formation \citep{kennicutt1998star, calzetti2007calibration, deMellos2024}. Also important are atomic (ionic) emission lines originating from heavy elements. These lines span a wide range of ionisation potentials (IPs), allowing different photon energies to be probed. Since each feedback mechanism interacts with and ionises the ISM differently, these metal lines originate from and serve as sensitive kinematic tracers across a wide range of ionisation energies, making them invaluable diagnostics for understanding and disentangling the roles of AGN and star-formation on the structure and dynamics at the centres of galaxies \citep{kewley2006host, ma2021spatially}.

Many atomic lines lie in the near- and mid-infrared (IR), a wavelength range ($\lambda_{\text{rest}} \sim 1-30 \mu$m) challenging to observe from the ground due to strong atmospheric absorption and high thermal background. Resolving the complex kinematics of the nuclear regions of galaxies requires high spatial resolution (typically tens of parsecs or better), making the JWST NIRSpec \citep{Jakobsen_2022, boker2023orbit} and MIRI/MRS \citep{wells2015mid, argyriou2023jwst} integral field spectrographs (IFS) exceptionally well-suited for such work, especially for nearby galaxies.

In addition to being characterised by their IP, many emission lines in the IR have been shown to trace specific physical components or processes within galaxies, particularly on nuclear scales. For example, [Fe II] 5.34$\mu$m is widely used as a tracer of shocked gas, especially when its flux ratio is high relative to hydrogen recombination lines \citep{kawara1988forbidden, colina2015understanding, vivian2022goals, herrero2025miconic}. Similarly, [Mg IV] 4.49$\mu$m has been shown to trace direct AGN photoionisation as well as faster shocks originating from more highly ionised regions \citep{pereira2024extended}. Several IR lines also serve as diagnostics of recent star-formation, including [Ar II] 6.99$\mu$m and [Ne II] 12.81$\mu$m \citep{ho2007mid, zhuang2019new, whitcomb2020comparative, young2023halfway}, while higher ionisation neon lines in the mid-IR, such as [Ne III] 15.56$\mu$m, [Ne V] 14.32$\mu$m, and [Ne VI] 7.65$\mu$m also provide a means to distinguish between different excitation mechanisms \citep{feltre2023optical, zhang2025theoretical}. Notably, [Ne II] and [Ne III] trace star-forming regions, with [Ne III] also being enhanced in more energetic or turbulent environments, whereas [Ne V] and [Ne VI], with significantly higher IPs, are exclusively associated with AGN activity \citep{sturm2002mid, pereira2010mid, munoz2025miconic}. Additional mid-IR lines such as [S IV] 10.51$\mu$m and [O IV] 25.89$\mu$m are also strong tracers of AGN ionisation and can highlight features such as ionised outflows \citep{dicken2014spitzer}.

Numerous studies have utilised JWST/NIRSpec and MIRI/MRS integral field spectroscopy (IFS) to investigate the nuclear regions of nearby galaxies (e.g. \citealt{garcia2022high, armus2023goals, donnan2023obscured, herrero2024miconic, garcia2024structures, munoz2024biconical, ceci2025jwst, Marconcini2025miri, feuillet2025core, riffel2025impact, almeida2025jwst, veenema2025shock}).

Of particular relevance to this study, as they probe the kinematics of different ISM ionisation phases as a function of IP, \citet{davies2024gatos} studied the nucleus of NGC~5728 with MIRI/MRS, demonstrating that molecular gas is being depleted by outflows along an AGN-driven ionisation cone. They also showed that the kinematic major axis differs when traced by mid-IR [Ne II], [Ne III], [Ne V], and [Ne VI] lines, with the higher-IP lines exhibiting clear offsets relative to lower-IP lines due to the different position angles (PAs) of the ordered disk rotation and ionisation cone (this was also shown in several other Seyfert galaxies by \citealt{zhang2024galaxy}, and by \citealt{munoz2024biconical}). 

Also of particular interest to this study, \citet{ulivi2025jwst} analysed the inner $\sim 1$ kpc of Arp~220 using JWST NIRSpec IFS. By examining multiple near-IR emission lines and fitting multi-Gaussian profiles, they disentangled the kinematics of the ISM, showing two counter-rotating disks, each associated with one of the nuclei, as well as two distinct outflows, each driven by an active nucleus.

An important factor in resolving and interpreting kinematic signatures in galaxies is the viewing geometry, specifically, the galaxy's inclination and our line-of-sight to its morphological components (such as nuclear rings, disks, jets, and outflows). The prevalence and morphology of these features across the galaxy population remain uncertain, motivating the need for detailed case-by-case studies. One such example is NGC~7582, a galaxy with a high inclination ($i \sim 58^\circ$) and a morphologically complex nuclear region. As we will show, NGC~7582 exhibits a pronounced kinematic dichotomy between low- and high-ionisation circumnuclear gas, which is distinguishable in part due to our viewing angle.

The remainder of this paper is structured as follows. In Section~\ref{sec:target}, we review previous studies of NGC~7582. Section~\ref{sec:method} describes the data collection, reduction, and analysis methods. Section~\ref{sec:resultsanddiscussion} gives our combined results and discussion, with Section~\ref{sec:singlegauss} presenting results from single-Gaussian fitting to many mid-IR emission lines across the nuclear and circumnuclear regions, showing strong variation between disk and outflow tracing lines. Section~\ref{sec:doublegauss} extends this analysis to double-Gaussian fits and discusses the decoupling of the disk and outflow kinematics in three intermediate-IP lines. Then in Section~\ref{sec:diskmodelling}, we apply a physically motivated thin inclined disk model to the intermediate-IP lines to isolate the outflow kinematic signature, while in Section~\ref{sec:conemodelling} we instead apply a simple one-dimensional outflow model to isolate the disk rotation. Building on these, Section~\ref{sec:hybridmodel} introduces a hybrid approach that simultaneously models both disk and outflow components. In Section~\ref{sec:escape_velocity}, we consider the expulsion of gas in the outflow. Then in Section~\ref{sec:opening_angle_vs_IP}, we measure the opening angle of the front facing ionisation cone in order to constrain the morphology of the AGN dusty torus in NGC~7582. Finally, Section~\ref{sec:conclusions} summarises our findings.

\section{NGC 7582}
\label{sec:target}

\begin{figure*}
   \centering
   \includegraphics[width=\linewidth]{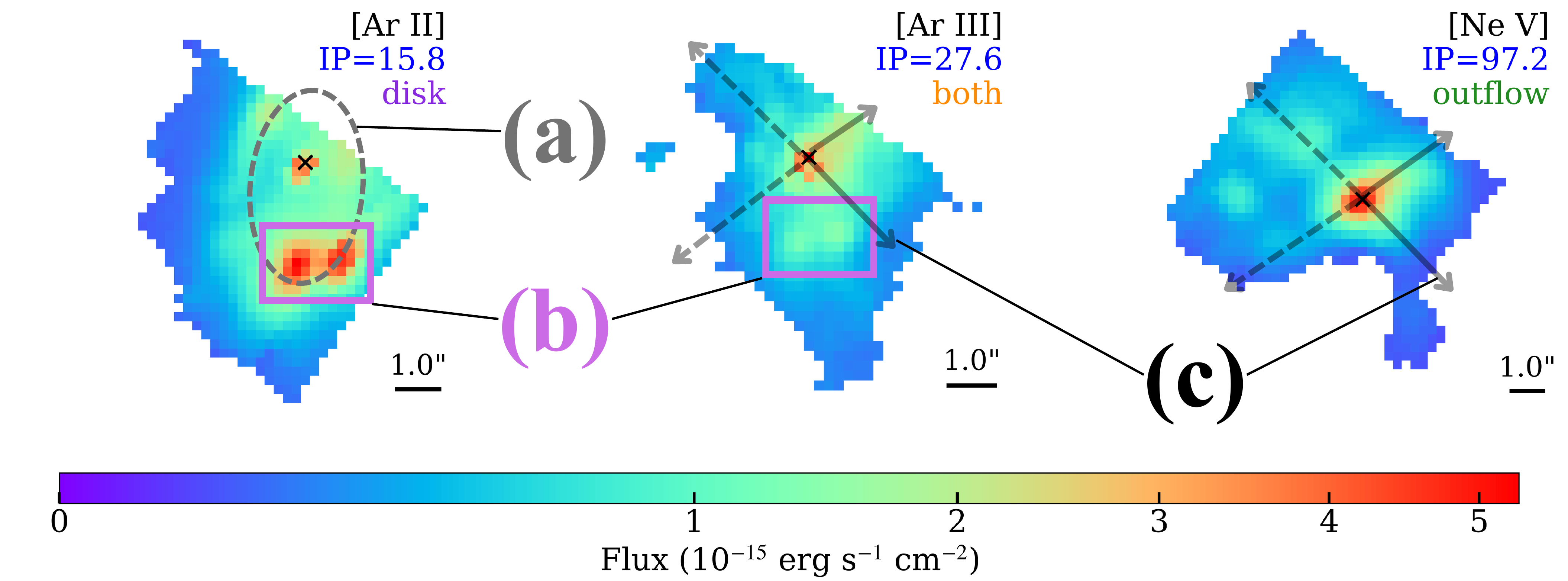}
   \caption{Integrated flux maps of the [Ar II] 15.8$\mu$m, [Ar III] 8.99$\mu$m, and [Ne V] 14.32$\mu$m emission lines, annotated to highlight the main nuclear and circumnuclear structures in NGC~7582. We give the line name, IP, and associated kinematic trace (see Table~\ref{tab:ion_lines}) at the top right of each flux map. North is up, east is to the left. The AGN position (defined as the peak continuum emission in each channel) is marked by a cross. (a) Outline of the circumnuclear disk/ring: a highly inclined ($i \sim 58^\circ$; \citealt{wold2006nuclear2, garcia2021galaxy}) structure rich in molecular gas and hosting multiple star-forming regions. This component is most clearly traced by the low-IP emission lines. (b) Southern circumnnuclear star-forming clumps: prominent regions embedded within the disk/ring, identified as SF~1 and SF~2 by \citet{veenema2025shock} (also referred to as M1 and M2 by \citealt{wold2006nuclear} and \citealt{ricci2018optical}). These clumps are bright in low- to intermediate-IP tracers. (c) Ionised biconical outflow: the AGN-driven, limb-brightened cone structure, with solid arrows indicating the front-facing (blueshifted, western) cone edges and dashed arrows marking the receding (eastern) side. The outflow is primarily traced by mid- and high-IP lines, appearing as diffuse extensions along the arrowed directions from the nucleus, with the western cone more prominently detected. As with Fig.~\ref{fig:flux_comp}, this figure illustrates how different emission lines preferentially trace distinct physical components of the galaxy, depending on their ionisation potential, and provides a visual guide to interpreting the kinematic and flux maps presented throughout this work.}
    \label{fig:schematic}
\end{figure*}

NGC~7582 is a nearby Seyfert 2 galaxy at redshift $z \sim 0.00525$, corresponding to a distance of $D \sim 22$~Mpc (assuming a flat $\Lambda$CDM Universe with $H_0 =70$ km/s/Mpc, $\Omega_m = 0.3$), therefore 1$\arcsec \sim 100$\:pc. Previous studies have identified several key nuclear and circumnuclear features, including a star-forming disk/ring at a radius of $\sim 200$~pc \citep{riffel2009agn, Alonso-Herrero2021, garcia2021galaxy, 2022ApJ...925..203J}, composed of clumps with elevated star-formation containing roughly $\sim 500$ O-type stars each, with a star formation rate of $0.23-0.28$ $M_\odot$/yr, \citealt{riffel2009agn}) to the north and south of a Compton-thick central AGN \citep{garcia2016nuclear, ricci2018optical}, with a hydrogen column density of $N_{\mathrm{H}} \sim 10^{23} - 10^{24} $ cm$^{-2}$ \citep{rivers2015nustar}. The galaxy hosts a prominent bar \citep{riffel2009agn}, and the $\sim 200$~pc radius circumnuclear ring is likely located near the inner Lindblad resonance, where bar-driven gas inflows accumulate and trigger the observed enhanced star formation \citep{sormani2024nuclear}.

Within $\sim 50$~pc of the nucleus, a marked depletion of molecular gas has been observed, while a dusty, molecular-rich torus is found on scales of $\gtrsim 10$~pc surrounding the AGN \citep{garcia2021galaxy, garcia2024deciphering}. Additionally, a pair of ionisation cones originating from the AGN and extending over $\gtrsim 3$ kpc has been confirmed, containing highly ionised, outflowing gas \citep{morris1985velocity, 2022ApJ...925..203J}.

In Fig.~\ref{fig:schematic} we present integrated flux maps (see Section~\ref{sec:method_calc} for details) of the mid-IR [Ar II] 15.8 $\mu$m, [Ar III] 8.99 $\mu$m, and [Ne V] 14.32 $\mu$m emission lines from JWST MIRI/MRS (see Section~\ref{sec:singlegauss}), with annotations highlighting the key morphological components of the nuclear and circumnuclear regions of NGC~7582. These include the circumnuclear star-forming disk/ring, the embedded southern star-forming clumps, the AGN position, and the edges of the biconical ionised outflow. The figure serves as a visual guide to the spatial distribution of these structures, which are traced to varying degrees across the full set of emission lines analysed in this work, and is intended to aid the interpretation of the flux and kinematic maps presented throughout the paper.

NGC~7582 has also been shown to have strong radio emission \citep{forbes1993radio, orienti2010radio} from its AGN out to $\sim4$${\arcsec}$ north and $\sim2$${\arcsec}$ south. This emission is rather diffuse, with current debate on whether it could be originated from intense starburst clumps in the circumnuclear disk, or a rather young radio jet (11,000-23,000 years old, \citealt{2022ApJ...925..203J}).

Of particular relevance is the work of \citet{ricci2018optical}, who analysed optical GMOS \citep{allington2002integral} and near-IR SINFONI \citep{eisenhauer2003sinfoni} observations of NGC~7582. They showed that the low-ionisation gas in the circumnuclear ring/disk, traced by H$\alpha$ and [N II] 6584\AA, follows the galactic disk rotation described by \citet{morris1985velocity}, with a major axis at PA $\sim 0 \pm 5^\circ$. In contrast, the high-ionisation [O III] 5007\AA~line traces the biconical ionised outflow, oriented at a different PA of $\sim 47 \pm 5^\circ$. \citet{ricci2018optical} also proposed a model of the circumnuclear structure of NGC~7582, comprising of the highly inclined circumnuclear ring surrounding the AGN torus, with the biconical outflow emerging from within the ring and oriented approximately orthogonal to it. They further incorporated the putative radio jet oriented nearly perpendicular to the ionisation cones (see their fig. 17).

Also noteworthy is the study by \citet{2022ApJ...925..203J}, who used VLT/MUSE \citep{bacon2010muse} observations of NGC~7582 to show that its circumnuclear star-forming ring/disk constitutes a kinematically distinct core (KDC), on a scale of $\sim 600$pc. The stars in the KDC corotate with the main disk of the galaxy along a similar kinematic axis (which they found had a PA $\sim -23^\circ$), but exhibit a much higher rotation velocity that rises steeply before peaking along the major axis. The low-ionisation gas kinematics in the ring/disk, traced by H$\alpha$ and [N II] 6584\AA, were found to closely follow the stellar component, rotating at the same PA and with similar velocity.

\citet{2022ApJ...925..203J} also examined the kinematics of the ionised conical outflow, again with [O III] 5007\AA. They found the outflow to have edges at PA $\sim 15^\circ$ and $\sim 115^\circ$, originating from the AGN torus and emerging through the circumnuclear ring/disk, consistent with ALMA observations by \citet{garcia2021galaxy}. Their analysis supports a biconical outflow with hollow ionisation cones, where enhanced emission along the cone edges arises from limb brightening, indicating that the cone interior contains more diffuse gas. They further identified evidence of gas at the cone edges moving more slowly as it interacts with the surrounding medium, possibly producing shocks. The redshifted side of the outflow shows weaker emission, attributed to obscuration by dust lanes.

These previous investigations of the gas kinematics in the nuclear regions of NGC~7582 \citep{ricci2018optical, 2022ApJ...925..203J} have primarily relied on optical and near-IR observations from ground-based facilities. While highly informative, such studies are limited by strong dust extinction at shorter wavelengths and by challenges inherent to ground-based observing, including atmospheric absorption and turbulence. By contrast, a mid-IR study is far less affected by dust obscuration.

In a previous study \citep{veenema2025shock}, we investigated the mid-IR molecular H$_2$ emission within the central $\sim 200$ pc of NGC~7582, finding that it follows the rotation of the circumnuclear disk/ring. We also showed evidence of heating by slow shocks ($v_s \sim 10$ km/s) in the southern clumps of the circumnuclear star-forming disk, likely driven by intense star-formation. Altogether, these earlier results highlight that the kinematics of the central $\sim$kpc of NGC~7582 are highly intricate and merit dedicated, detailed study.

\section{Method, Data collection and reduction}
\label{sec:method}

\subsection{Data collection and reduction}

The data used here are part of the Galaxy Activity, Torus, and Outflow Survey ({\textcolor{blue}{\href{https://gatos.myportfolio.com/}{GATOS}}}; \citealt{garcia2021galaxy, alonso2021galaxy}). These data are part of the JWST Cycle 2 GO proposal ID\,3535 (PIs: I. Garc\'ia-Bernete and D. Rigopoulou), which were originally presented and the data reduction discussed in \citet{veenema2025shock}. This study uses near-IR to mid-IR (2.87-28.1~$\mu$m) observations using integral-field spectrographs MIRI/MRS (4.9-27.9~$\mu$m) with a spectral resolution of R$\sim$1300-3700 (\citealt{labiano2021wavelength}) ($\sigma_{\text{inst}} \sim 35-97$ km/s), acquired on October 31st 2023, and NIRSpec with the grating-filter pairs G395H/F290LP (2.87-5.27~$\mu$m) with R$\sim$2700 (\citealt{Jakobsen_2022,boker2022near}) ($\sigma_{\text{inst}} \sim 47$ km/s), acquired on 7th April 2024.

\subsection{Calculating emission line flux and kinematics}
\label{sec:method_calc}

In this study we focus on the kinematics of various emission lines in the mid-IR. For each line, per spaxel, we first masked the spectral emission region and fitted a third-order polynomial out to 0.02$\mu$m of the continuum on either side. This continuum fit was then subtracted from all wavelength bins to isolate the emission line.

We then masked out spaxels which had a line peak signal-to-noise (S/N) less than 3, and those with emission line fluxes that were lower than the 60th percentile of the flux of that specific emission line across every spaxel. This approach retains the brightest 40\% of spaxels for each line, preferentially selecting regions with sufficient flux for reliable Gaussian kinematic fitting. This gives a variable S/N cut for each emission line, with the minimum S/N (i.e. the S/N at the 60th flux percentile) being > 6, but typically much higher, for all emission lines. This additional flux threshold is important because the velocity and velocity dispersion maps are not flux-weighted, and without such a cut, low-flux, or more noisy spaxels would contribute equally to the kinematic analyses presented here. This threshold therefore balances statistical robustness and spatial coverage: lower cuts admit more noise-dominated spaxels, while higher cuts overly restrict the FoV and reduce the number of spaxels available for analysis.

For each emission line in every remaining (unmasked) spaxel, we fit a single-Gaussian profile, with the best-fit central wavelength ($\lambda$) used to derive the gas velocity,  $v = c \times(\lambda-\lambda_0)/\lambda_0$, where $c$ is the speed of light and $\lambda_0$ is the rest-frame wavelength of the emission line. We also attempted double-Gaussian fits across all lines. However, we found that for many emission lines, double-Gaussian functions did not statistically improve the fit compared to a single-Gaussian function. These double-Gaussian fits were implemented under various model assumptions, discussed in detail later. All single- and double-Gaussian fitting was performed using Levenberg-Marquardt (LM) optimisation via the CapFit Python library \citep{cappellari2023full}. We show several example individual spaxel fits in the Appendix in Figs.~\ref{fig:ArII_spaxel}, \ref{fig:SIV_spaxel}, \ref{fig:NeIII_spaxel}.

\begin{figure*}
   \centering
   \includegraphics[width=\linewidth]{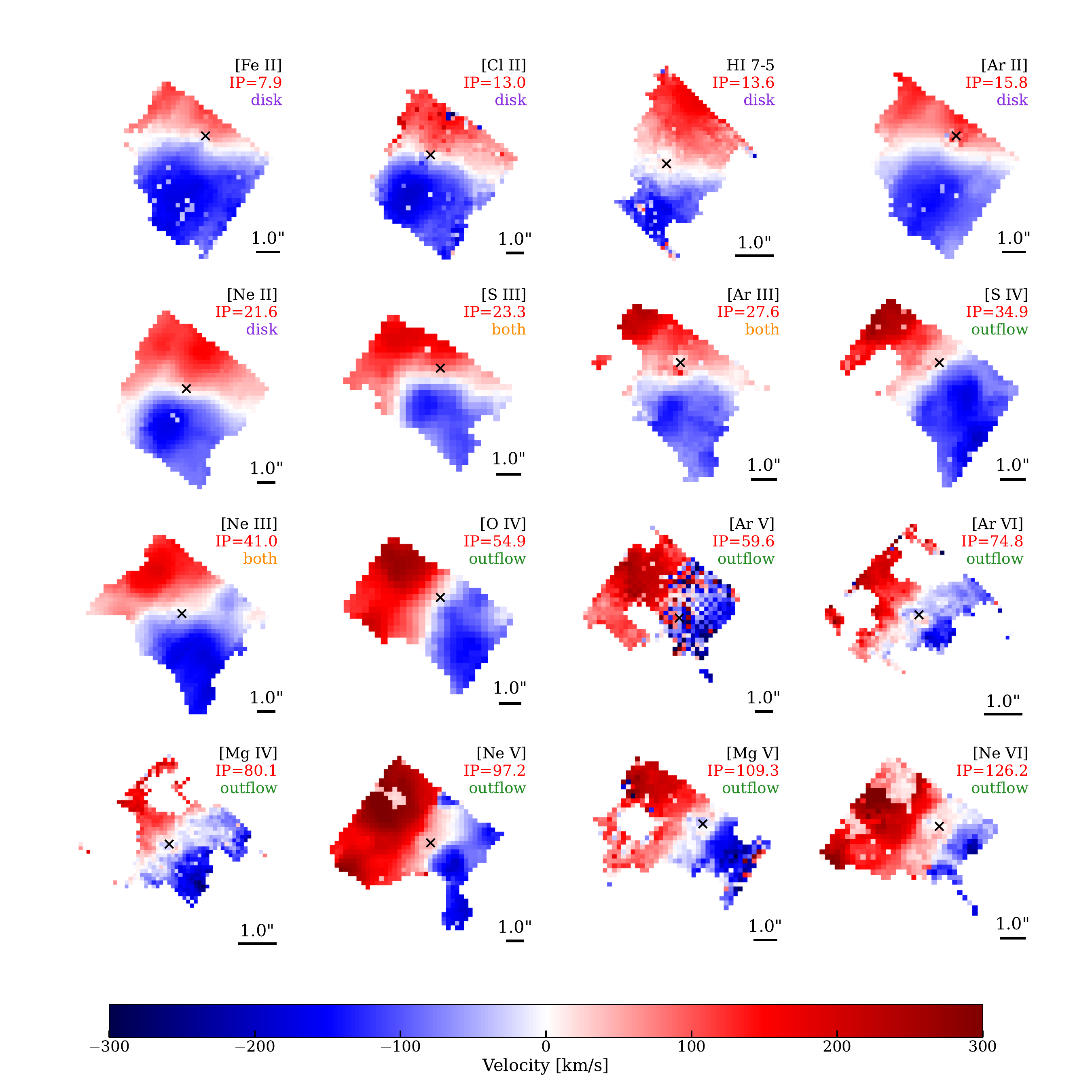}
   \caption{Velocity maps in the central $\sim 200$ pc region for various atomic emission lines in NGC~7582 fit using single-Gaussians ordered by IP (in eV) from left to right, top to bottom. Labels at the top right give the line name, IP, and kinematic trace for each velocity map. The lower IP lines typically trace the ordered circumnuclear star-forming ring/disk rotation, whereas the higher IP lines typically trace the biconical ionised outflow motion. North is up, east is to the left. Each black cross denotes the AGN position, i.e. the photometric centre of the corresponding continuum. A 1$\arcsec \sim 100$ pc scalebar is included at the bottom right of each subplot.}
    \label{fig:single_gauss_vel_comp}%
\end{figure*}

\begin{figure*}
   \centering
   \includegraphics[width=\linewidth]{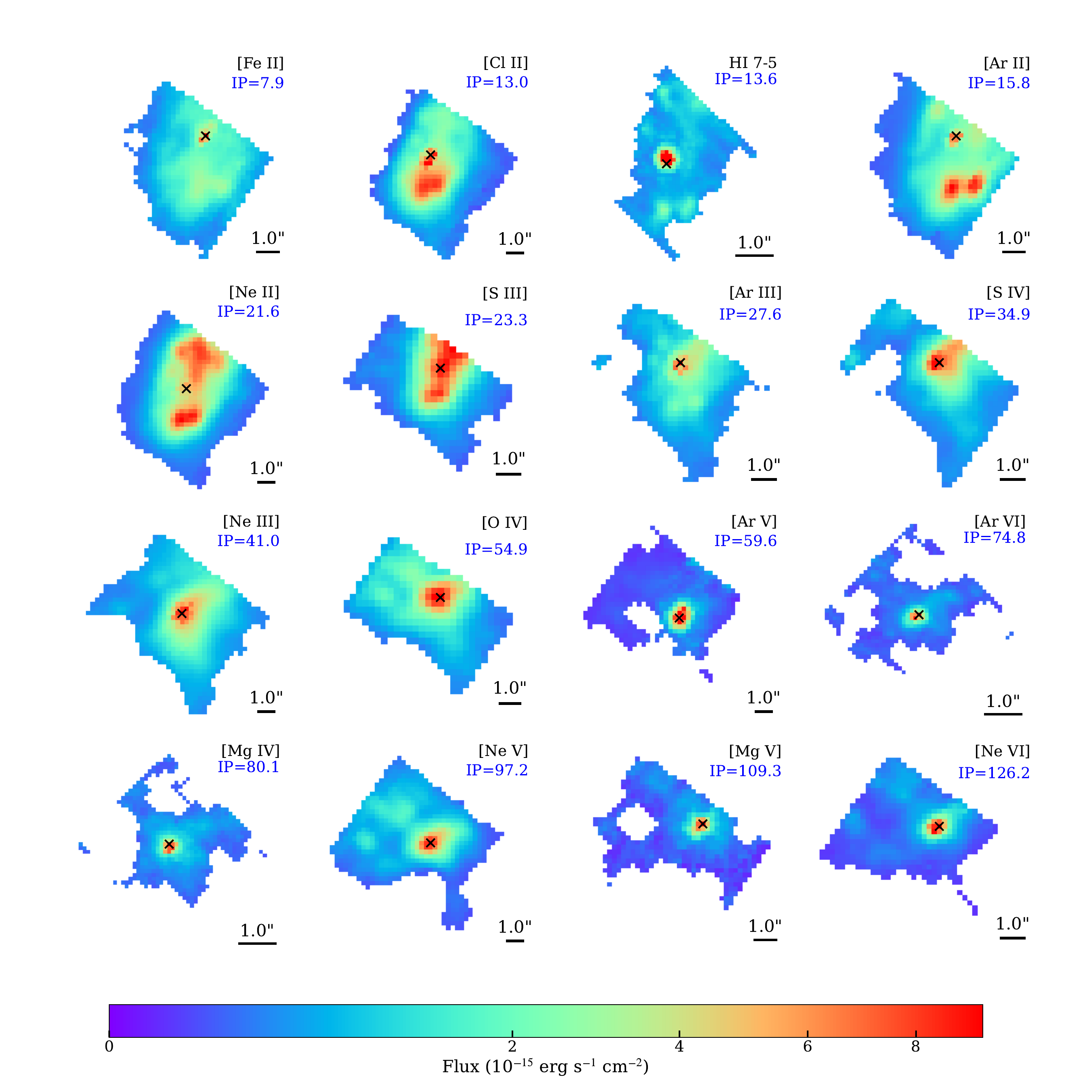}
   \caption{Continuum subtracted flux maps in the central $\sim 200$ pc region for various atomic emission lines in NGC~7582 ordered by IP (in eV) from left to right, top to bottom. The lower IP lines typically trace the circumnuclear star-forming ring/disk, with multiple prominent clumps of star-formation visible, whereas the higher IP lines typically trace gas excited by the biconical ionised outflow. North is up, east is to the left. Each black cross denotes the AGN position, i.e. the photometric centre of the corresponding continuum. A 1$\arcsec \sim 100$ pc scalebar is included at the bottom right of each subplot.}
    \label{fig:flux_comp}%
\end{figure*}

\begin{figure*}
   \centering
   \includegraphics[width=\linewidth]{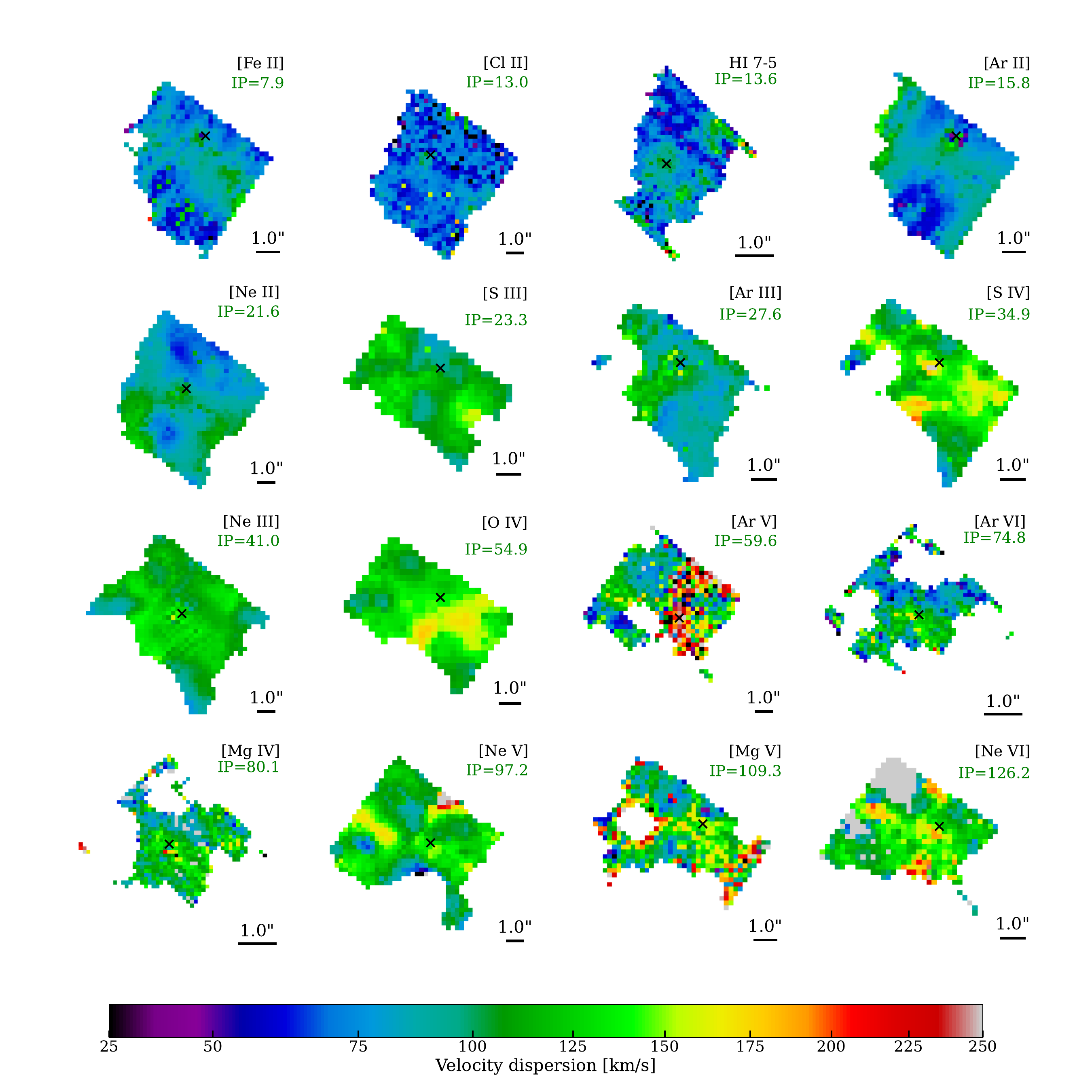}
   \caption{Velocity dispersion maps in the central $\sim 200$ pc region for various atomic emission lines in NGC~7582 ordered by IP (in eV) from left to right, top to bottom. These maps include the contribution from instrumental dispersion. The lower IP lines show lower velocity dispersions throughout the FoV, especially over the star-forming ring. The higher IP lines tend to have much larger velocity dispersions, particularly in the biconical outflow. Each black cross denotes the AGN position, i.e. the photometric centre of the corresponding continuum. A 1$\arcsec \sim 100$ pc scalebar is included at the bottom right of each subplot.}
    \label{fig:disp_comp}%
\end{figure*}

We generated continuum subtracted integrated flux emission and velocity maps for each emission line using the native resolution and field of view (FoV) of the NIRSpec or MIRI/MRS datacubes, without reprojection or convolution. In this study, we are analysing many different emission lines that originate from different instruments (NIRSpec or MIRI/MRS) or MIRI/MRS channels, each with slightly different spaxel scales, resolutions, and FoVs. Hence, the spatial sampling and exact coverage vary slightly between lines. For clarity, and to prevent overcrowding in multi-panel figures, we omit explicit RA and Dec axes in most figures, however we do mark the central AGN position (as the position of the peak continuum flux in each channel) on every map as a point of reference and include scalebars on our main reference figures showing the single-Gaussian velocity, flux, and single-Gaussian velocity dispersion maps, Figs.~\ref{fig:single_gauss_vel_comp},~\ref{fig:flux_comp},~\ref{fig:disp_comp}.

\section{Results and discussion}
\label{sec:resultsanddiscussion}

\begin{table*}
    \centering
    \caption[]{Key fine structure transitions analysed in this study, including their ionisation potential (IP) (defined as the energy needed to reach the ionisation stage producing that transition, \citealt{kramidateam}), rest frame wavelength, the associated kinematic component traced, major velocity axis position angle (from PAFit), and median velocity dispersion (from single-Gaussian fitting).}
    \label{tab:ion_lines}
    \begin{tabular}{lccccc}
        \hline
        Emission Line & IP (eV) & Wavelength ($\mu$m) & Kinematic Trace & Kinematic major axis PA (degrees) & Average velocity dispersion ($\sigma$) (km/s) \\
        \hline
        {[Fe II]} & 7.9   & 5.34  & Disk          & -12 $\pm$ 13 & 76 $\pm$ 10 \\
        {[Cl II]} & 13.0  & 14.37 & Disk          & -11 $\pm$ 9 & 64 $\pm$ 17 \\
        HI 7-5    & 13.6  & 4.65  & Disk          & -13 $\pm$ 8 & 71 $\pm$ 17\\
        {[Ar II]} & 15.8  & 6.99  & Disk          & -17 $\pm$ 19 & 85 $\pm$ 9 \\
        {[Ne II]} & 21.6  & 12.81 & Disk          & -9 $\pm$ 11 & 92 $\pm 13$ \\
        {[S III]} & 23.3  & 18.71 & Disk/Outflow  & 13 $\pm$ 11 & 115 $\pm$ 14 \\
        {[Ar III]}& 27.6  & 8.99  & Disk/Outflow  & 13 $\pm$ 13 & 94 $\pm$ 13  \\
        {[S IV]}  & 34.9  & 10.51 & Outflow       & 58 $\pm$ 14 & 129 $\pm$ 17  \\
        {[Ne III]}& 41.0  & 15.56 & Disk/Outflow  & 17 $\pm$ 9 & 115 $\pm$ 11  \\
        {[O IV]}  & 54.9  & 25.89 & Outflow       & 42 $\pm$ 12 & 131 $\pm$ 10 \\
        {[Ar V]}  & 59.6  & 13.10 & Outflow       & 62 $\pm$ 2 & 115 $\pm$ 31  \\
        {[Ar VI]} & 74.8  & 4.53  & Outflow       & 56 $\pm$ 8 & 92 $\pm$ 23 \\
        {[Mg IV]} & 80.1  & 4.49  & Outflow       & 62 $\pm$ 2 & 115 $\pm$ 25  \\
        {[Ne V]}  & 97.2  & 14.32 & Outflow       & 65 $\pm$ 10 & 115 $\pm$ 22  \\
        {[Mg V]}  & 109.3 & 5.61  & Outflow       & 43 $\pm$ 9 & 124 $\pm$ 32 \\
        {[Ne VI]} & 126.2 & 7.65  & Outflow       & 44 $\pm$ 12 & 131 $\pm$ 26  \\
        \hline
    \end{tabular}
\end{table*}

\subsection{Single-Gaussian fitting}
\label{sec:singlegauss}

We begin by broadly examining the kinematics of several emission lines detected across the combined NIRSpec and MIRI/MRS wavelength range for NGC~7582. These lines, along with their wavelengths, IPs, interpreted kinematic trace, major velocity axis position angle, and average velocity dispersion ($\sigma$) from single-Gaussian fitting are listed in Table~\ref{tab:ion_lines}. Fig.~\ref{fig:single_gauss_vel_comp} presents the velocity maps from the single-Gaussian fits to the continuum-subtracted flux for each line, ordered by increasing IP. We also show the continuum subtracted line flux emission for each emission line in Fig.~\ref{fig:flux_comp}, as well as the velocity dispersion in Fig.~\ref{fig:disp_comp}. 

Notably, when fitting single-Gaussian profiles to the emission lines and deriving the corresponding velocity maps, we observe a striking kinematic dichotomy between low-IP ($\lesssim 20$ eV) and high-IP ($\gtrsim 35$ eV) lines, as shown in Fig.~\ref{fig:single_gauss_vel_comp}. The low-IP lines exhibit a rotation pattern with blueshifted emission to the south of the AGN and redshifted emission to the north. In contrast, high-IP lines display a different kinematic axis, with blueshifted emission to the southwest and redshifted to the northeast.

Additionally, we analysed the kinematics of the pure rotational H$_2$ lines from S(1) to S(7), which trace the warm molecular gas. These lines all exhibit the same strongly disk-like rotation as the low-IP atomic lines. Several of these line kinematics are presented and discussed in detail in \citet{veenema2025shock}, and \citet{donnan2026}.

\subsubsection{Low- and high-IP lines: single-Gaussian}

The kinematics of low-IP lines in the circumnuclear region in the optical and near-IR have been discussed in previous studies \citep{morris1985velocity, riffel2009agn, ricci2018optical}, which attributed the observed kinematics to rotation within the actively star-forming circumnuclear ring/disk. We too see evidence for this in our flux maps, Fig.~\ref{fig:flux_comp}, with many showing strong flux from the southern two prominent star-forming clumps (seen as the bright spots roughly 1$\arcsec$ south of the AGN, particularly visible for [Ar II]), and some of the northern clumps (when resolved and within the FoV).

The high-IP line flux maps (Fig.~\ref{fig:flux_comp}) are dominated by compact emission from the nuclear region, consistent with photoionisation by the AGN as the primary source of these lines. This is particularly evident for the highest-IP species (IP $\gtrsim 55$ eV), which are largely confined to the central spaxels. However, we also identify visual evidence for extended emission in many of these high-IP line flux maps. This is most notable in [Ne V], [S IV], [Ar VI], and [Mg IV], which exhibit faint but coherent structure extending westward from the nucleus. This extension spatially coincides with the blueshifted side of the ionised outflow (Fig.~\ref{fig:schematic}) and displays a more diffuse morphology than the compact nuclear emission, and is broadly consistent with a hollow-cone geometry. We therefore interpret the bulk of the high-IP emission as tracing AGN photoionisation in the nucleus, while also noting evidence for an additional, spatially extended component likely associated with the outflow. The ionisation cone is also slightly traced in the higher-IP line velocity dispersion maps (Fig.~\ref{fig:disp_comp}) as regions of significantly enhanced dispersion (i.e. the more yellow and red regions at a PA $\sim 45^\circ$).

\begin{figure*}
   \centering
   \includegraphics[width=0.49\linewidth]{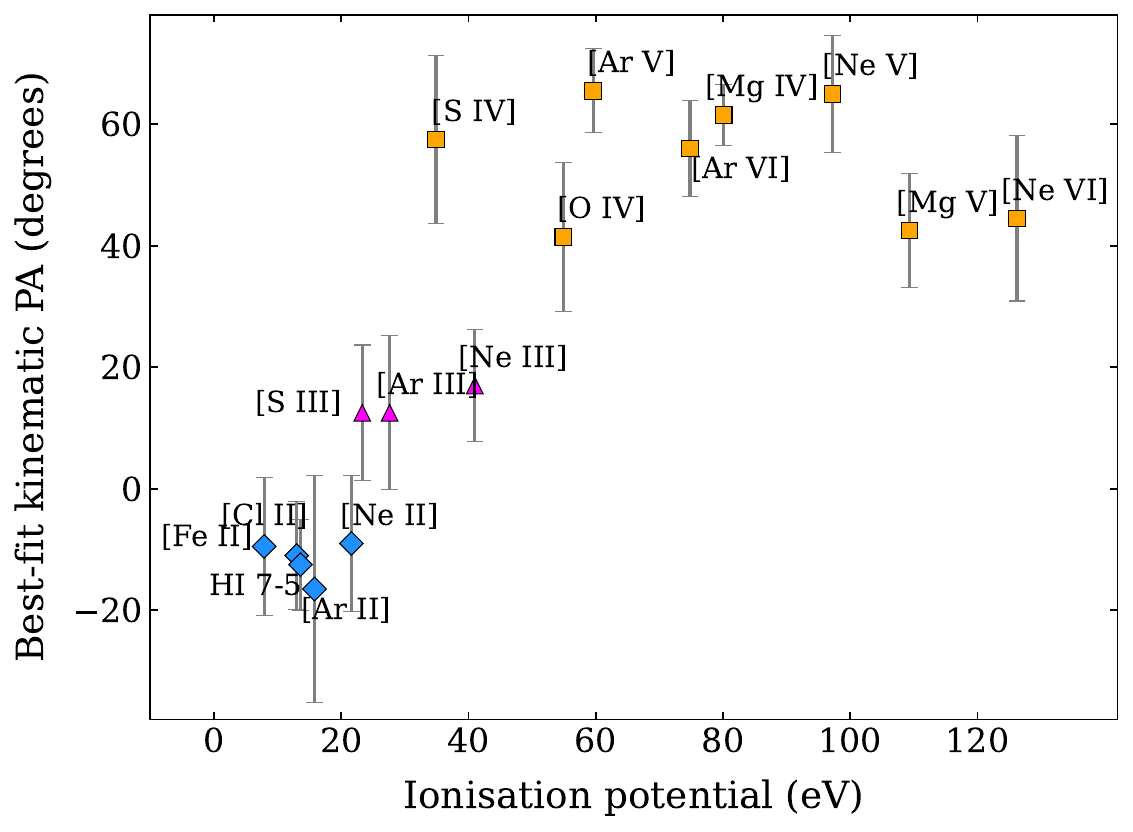}
   \includegraphics[width=0.49\linewidth]{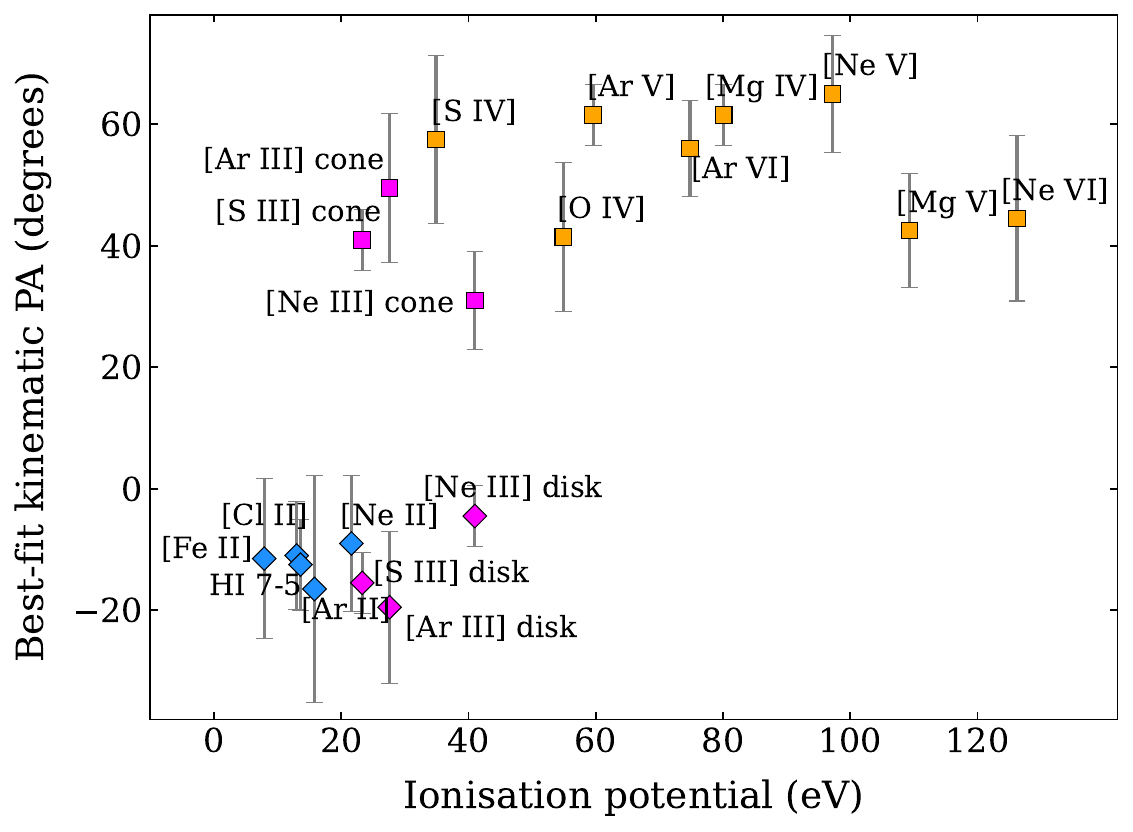}
   \caption{Left: best fit major kinematic axis position angle, (PA) vs ionisation potential (IP) for each of the lines from the single-Gaussian velocity maps fit using PAFit \citep{Krajnovic2006}. Errors are the 1$\sigma$ errors. Blue diamond points are emission lines we denote as disk-tracing low-IP, magenta triangle points denote the mixed kinematics tracing `intermediate-IP' lines, and orange square points denote outflow-tracing high-IP lines. The blue diamond points have an average PA $\sim -12 \pm 3^\circ$ whereas the orange square points have an average PA $\sim 54 \pm 10^\circ$. The magenta triangle points lie in the transition region between the other two groups and have an average PA $\sim 14\pm2^\circ$. Right: same as the left but with the intermediate-IP line PAs split into their disk component and outflow component velocity map PAs from double-Gaussian fitting (Fig.~\ref{fig:double_gauss_vel_comp}).}
    \label{fig:PA_vs_IP}%
\end{figure*}

Moreover, \citet{2022ApJ...925..203J} found the low-IP gas rotates with a kinematic major axis PA $\sim -23^\circ$ (in our notation going from red to blue shifting), while the high-IP [O III] line exhibited a PA $\sim 57^\circ$, with \citet{ricci2018optical} reporting similar PAs ($\sim 0^\circ$ for the low-IP and $\sim 47^\circ$ for the high-IP). Additionally, \citet{garcia2021galaxy} found that the cold molecular gas traced through CO 3-2 kinematics followed the disk rotation with a PA $\sim -16^\circ$. We show this is consistent within uncertainties with the results we observe in the mid-IR by fitting for the kinematic major axis PA using PAFit \citep{Krajnovic2006} on all single-Gaussian velocity maps in Fig.~\ref{fig:single_gauss_vel_comp}. We show this as a function of IP in Fig.~\ref{fig:PA_vs_IP} (left) and tabulate in Table~\ref{tab:ion_lines}. We estimate the kinematic PA uncertainty from the $\chi^{2}$ minimisation curve (from PAFit) as the angular range over which $\chi^{2}$ increases by one from its minimum (i.e. 1$\sigma$ confidence), weighted by the velocity uncertainties.

Furthermore, \citet{2022ApJ...925..203J} measured the ionised gas kinematics in the inner region using optical emission lines, finding that low-IP tracers such as H$\alpha$ closely follow the stellar disk rotation, with maximum rotational velocities of $\sim 150$ km/s. In contrast, the high-IP [O III] emission associated with the outflow reaches projected velocities up to $\sim 300$ km/s. Our mid-IR measurements show broad agreement: the low-IP lines analysed here exhibit maximum velocities of $\sim$ 150–180 km/s (e.g., see Table~\ref{tab:vmax_rturn}), while the high-IP lines reach comparable velocities along the ionisation cone. Together with the agreement in kinematic major axis PAs described above, this comparison supports the interpretation that the low- and high-IP mid-IR lines trace the same components previously identified in the optical, providing a consistent picture across wavelength regimes.

Therefore, based on previous findings and physical interpretation, we propose that the low-IP lines are tracing gas rotating within the circumnuclear star-forming disk, whereas the high-IP lines are tracing gas in the ionised biconical outflow. This classification is both physically motivated based on the emission and kinematic morphology, and supported by prior literature. Henceforth, we adopt the terms `disk-tracing' and `cone/outflow-tracing' as shorthand to describe the kinematic behaviour of respective emission lines.

We see that the disk-tracing lines, [Fe II], [Cl II], HI 7-5, [Ar II], and [Ne II], all exhibit nearly identical PAs, consistent with an average major axis PA~$\sim -12 \pm 3^\circ$. The outflow-tracing lines, [S IV], [O IV], [Ar V], [Ar VI], [Mg IV], [Ne V], [Mg V], and [Ne VI] also exhibit similar PAs, albeit a different PA to the disk tracing lines, consistent with an average major axis PA~$\sim 54 \pm 10^\circ$.

The masked flux maps also show the morphology of the disk and ionised outflow. Because only the brightest 40\% of spaxels are retained (Figs.~\ref{fig:single_gauss_vel_comp}, \ref{fig:flux_comp}, \ref{fig:disp_comp}), the spatial extent of each map directly traces the distribution of the corresponding emission line. Disk-tracing lines exhibit a roughly symmetric extension around the AGN, consistent with rotation about a central kinematic axis, while outflow-tracing lines extend predominantly to the north-east and south-west, in alignment along the ionisation cone axis.

\subsubsection{Intermediate-IP lines: single-Gaussian}

Of particular interest are what we will refer to as `intermediate-IP' lines, namely [S III], [Ar III], and [Ne III], which exhibit mixed kinematic signatures in the single-Gaussian velocity maps, Fig.~\ref{fig:single_gauss_vel_comp}. Some of these lines display regions which align more closely with the disk and others more with the outflow kinematics. We interpret this mixed behaviour as a natural consequence of their intermediate ionisation potentials: these lines can be produced both in the lower-ionisation, star-forming disk and in the more highly ionised outflowing gas. This behaviour likely reflects overlapping gas elements along the line of sight, where gas excited by both components (the disk and the outflow) is capable of producing these emission lines. Thus, it is unsurprising that they exhibit features of both the disk and outflow kinematic signatures. We also see evidence for this mixed behaviour in the flux maps, Fig.~\ref{fig:flux_comp}, where these lines appear to show traces of the disk, and the edges of the blueshifted ionisation cone. This is particularly visible in the [Ar III] emission map, where the cone edges appear as diffuse extensions westwards of the AGN, while we can also still resolve the two southern star-forming clumps, confirming that this line is able to trace both the outflow and the disk. This is also reflected in the best-fit major axis PAs, Fig.~\ref{fig:PA_vs_IP} (left). The [S III], [Ar III], and [Ne III] lines do not align with either the disk-tracing lines at $\sim -12^\circ$, IP $\lesssim 20$ eV, or the outflow-tracing lines at $\sim 54^\circ$, IP $\gtrsim 35$ eV, but instead fall between with an average PA $\sim 14 \pm 2^\circ$.

Also noteworthy is the contrasting behaviour of [S IV] and [Ne III]. While [S IV] (IP = 34.8 eV) appears to exclusively trace the outflow in both flux and velocity, [Ne III] (IP = 41.0 eV) shows contributions from both disk and outflow components. This demonstrates that IP alone does not fully determine which component (disk or outflow) the line traces. Rather, IP acts as a strong, but not exclusive determinant. The clean outflow signature in [S IV], despite its moderate IP, suggests it is preferentially excited by ionisation mechanisms active in the outflow rather than in the disk. Previous studies have shown that the [S IV] line traces both star formation and AGN activity across many active galaxies \citep{pereira2010local, esparza2018circumnuclear}. In some systems it primarily follows star-forming regions, in others the AGN, and in some it reflects a combination of both. In NGC~7582, we find that [S IV] traces mostly the AGN and its ionised outflow, with no indication of contribution from the star-forming disk in either its velocity or flux maps.

A similar trend was reported in other Seyfert galaxies by \citet{zhang2024galaxy}, who attributed the uniquely weaker detection of disk rotation in the [S IV] kinematics to obscuration, as this line lies near the centre of the 9.7 $\mu$m silicate absorption feature. Consequently, [S IV] emission from the star-forming disk is preferentially attenuated by silicate absorption, whereas emission from the ionised outflow, being located above and below the disk plane and thus subject to lower dust columns, remains largely unobscured. This naturally explains the weak or absent disk rotation signature in the [S IV] kinematics, despite the line's intrinsic sensitivity to both star formation and AGN excitation based on IP.

It is also interesting to note that the separation in IP between the lines we classify as low-IP lies close to the IP of neutral He (24.6 eV), while the division between the high-IP lines, from [O IV] upwards, occurs near the IP of He II (54.4 eV). This is not a coincidence, with studies such as \citet{feuillet2025core} likewise showing that ionic species capable of tracing star-formation predominantly have IPs below the He threshold, whereas those that trace AGN activity require IPs above the He II threshold. This cutoff is physically well motivated. Helium is the second most abundant element in the ISM, and its large photoionisation cross-sections at 24.6 eV and 54.4 eV act as efficient photon sinks. In star-forming regions, the UV spectra of massive stars already decline steeply with increasing energy; any photons above 24.6 eV are quickly photoabsorbed by He, and the scarcity of stellar photons beyond 54.4 eV means that He II effectively absorbs almost all of them. Consequently, star-forming regions, such as the circumnuclear disk, produce very few photons capable of creating ions with IPs above the He II edge. By contrast, AGN generate a hard, power-law continuum extending well into the extreme-UV and X-ray, easily supplying photons energetic enough to ionise species with IPs $\gtrsim $50-100 eV. Thus, ions with IPs below 24.6 eV can be maintained in stellar photoionised gas, whereas species with IPs exceeding 54.4 eV necessarily trace the much harder radiation fields associated with AGN.

\subsubsection{Velocity dispersion: single-Gaussian}

Similar IP-based behaviour is observed across the velocity dispersion maps of the low-, intermediate-, and high-IP lines (Fig.~\ref{fig:disp_comp}). These velocity dispersion maps look ``noisy'', not due to low S/N, but because of the limitations of fitting a single-Gaussian to a line profile that arises from two distinct kinematic components (see Section~\ref{sec:doublegauss} for physically motivated double-Gaussian fits).
% In particular, 
The low-IP lines exhibit the lowest velocity dispersions, with only modest enhancements near the AGN. The star-forming clumps north and south of the AGN, previously identified as part of the circumnuclear star-forming disk \citep{riffel2009agn, ricci2018optical, veenema2025shock}, roughly coincide with regions of reduced velocity dispersion in many low-IP lines, most notable in [Ne II], when compared with their flux distributions (Fig.~\ref{fig:flux_comp}). While our data alone do not directly demonstrate that these clumps are rotationally supported, the systematically lower dispersion observed in these more strongly emitting regions is consistent with dynamically colder gas, as expected for material embedded within a rotating disk where ordered motions dominate over local velocity dispersion.

The high-IP lines in Fig.~\ref{fig:disp_comp} show broadly coherent behaviour, with systematically higher velocity dispersions across the FoV compared to the low-IP lines, and indications of modestly enhanced dispersion aligned with the ionisation cone axis. Hence, the elevated dispersion, particularly along PA $\sim 45^\circ$ from the AGN, may be consistent with kinematic signatures of outflowing gas. This interpretation is supported by the results of \citet{2022ApJ...925..203J}, who analysed the velocity dispersion of ionised gas in the bicone (their figure~6e) and found enhanced values within, and especially along the edges of, the front-facing cone. The reported dispersions of $\sim 140\text{-}200$ km/s are comparable to the elevated values observed here in the high-IP lines. This agreement suggests consistency between the optical and mid-IR tracers, implying that the front facing ionisation cone is not significantly affected by dust extinction, and supports the interpretation that the regions of increased dispersion are associated with the ionised outflow.

The intermediate-IP lines exhibit a distinct mixed behaviour in the velocity dispersion maps. Specifically, they show reduced dispersion in the north-south direction associated with the star-forming clumps, similar to the low-IP lines, alongside moderately enhanced dispersion along the north-east to south-west axis coincident with the ionisation cone, as seen in the high-IP lines. This behaviour is particularly evident in [S III], where regions of elevated dispersion (yellow) to the top left and bottom right align with the ionisation cone axis, while lower-dispersion regions (darker green) north and south of the AGN trace the star-forming disk. This indicates that the intermediate-IP lines effectively capture the transition between the kinematic regimes traced by the low- and high-IP lines. Overall, the velocity dispersion maps reveal signatures of both the disk and outflow components, allowing a direct spatial comparison between the two. They further demonstrate that the intermediate-IP lines trace both kinematic structures simultaneously.

\begin{figure*}
   \centering
   \includegraphics[width=0.49\linewidth]{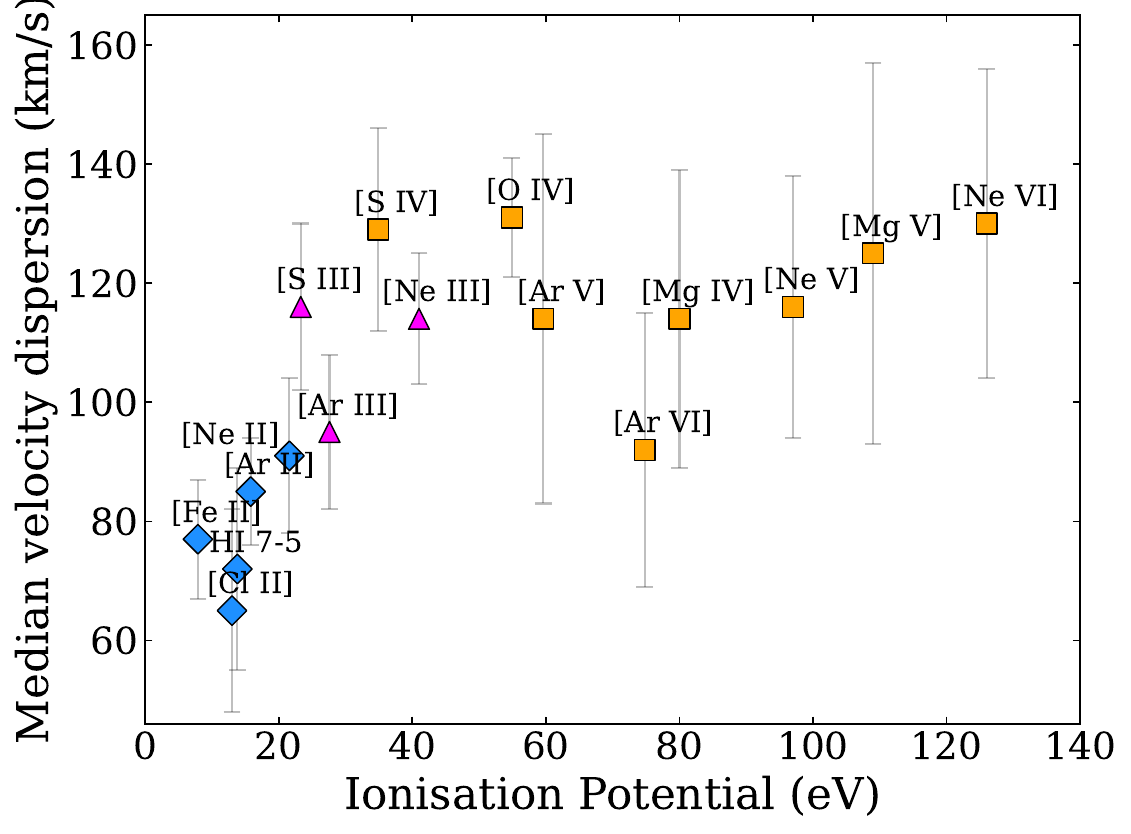}
   \includegraphics[width=0.49\linewidth]{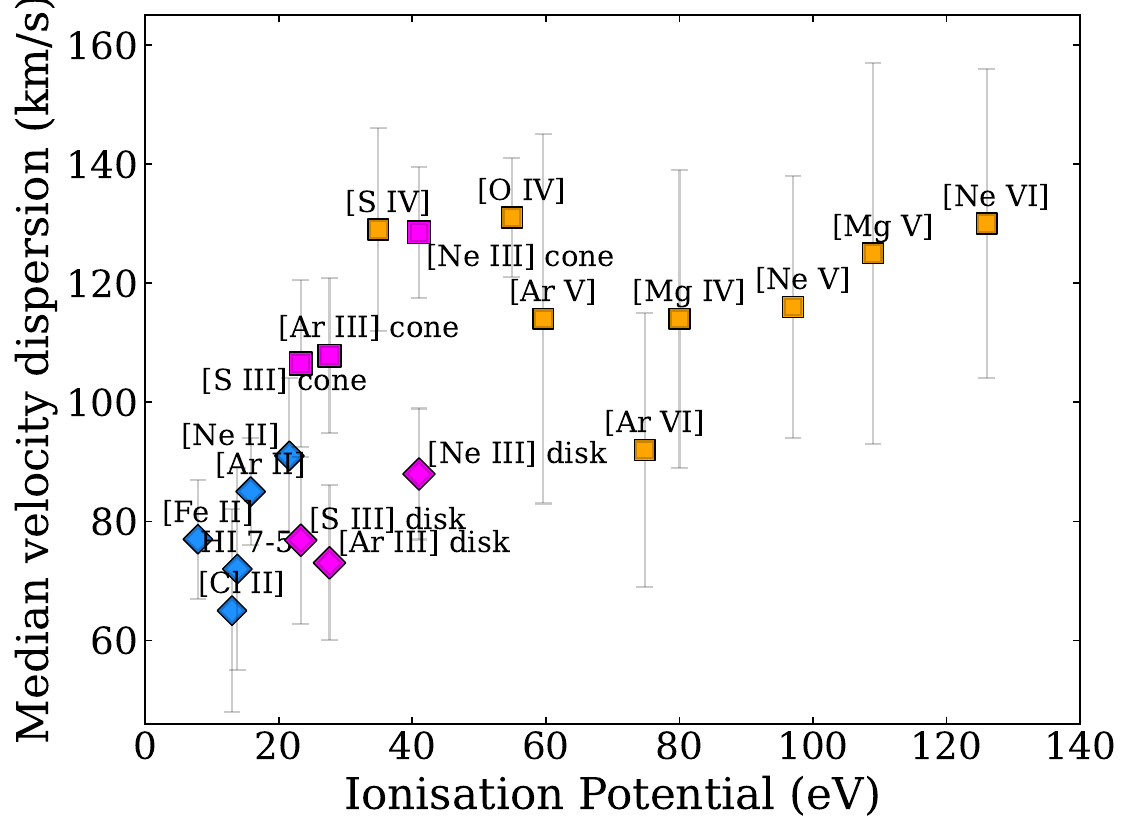}
   \caption{Left: median velocity dispersion of each emission line as a function of IP. We postulate that the blue diamond points lie roughly on a positive trend, whereas the orange square points display an approximately positive correlation within their errors. The magenta triangle points are in the transition region between the two distinct groups. Right: median velocity dispersion of each emission line, but with the intermediate-IP lines having both their median disk and outflow velocity dispersion found from double-Gaussian fitting (i.e. the dotted lines for each distribution shown in Fig.~\ref{fig:double_gauss_dispersion_histograms}) plotted instead.}
    \label{fig:mean_and_median_dispersions}%
\end{figure*}

We also compute the median velocity dispersion of each emission line as a function of IP across all unmasked spaxels, shown in Fig.~\ref{fig:mean_and_median_dispersions} (left), and Table~\ref{tab:ion_lines}. Both the median and mean (not plotted) dispersions exhibit a positive correlation with IP. This relation appears approximately linear for the low-IP, mostly disk-tracing lines (shown as blue diamond points). High-IP lines (shown as orange square points) show significantly higher velocity dispersions, and follow a similar or even flatter trend with IP but with significantly greater scatter. Similar velocity dispersion-IP trends have also been reported for other galaxies e.g., \citet{dasyra2011view} using archival Spitzer data, and \citet{munoz2024biconical} with MIRI/MRS.

The intermediate-IP lines (shown as magenta triangles) fall in between, with [S III] and [Ne III] both having relatively high median velocity dispersions consistent with the outflow-tracing lines, whereas [Ar III] shows a significantly lower median velocity dispersion, more along the trend of the disk-tracing lines. These three lines show hybrid behaviour with the positions of their points lying roughly in between the disk and outflow tracing values, once more suggesting that they are being efficiently excited in both circumnuclear structures. 

The contrasting trends in the velocity dispersion-IP relation highlight the different kinematic environments of the disk and the outflow. In the disk, the roughly linear increase of velocity dispersion with IP may arise because lower-IP (up to $\lesssim 20$ eV) emission preferentially traces star-forming clumps, which are present in the disk of this galaxy \citep{veenema2025shock}, where enhanced turbulence and local velocity gradients broaden the line profiles, naturally producing the observed trend. In contrast, the nearly constant high velocity dispersion of the high-IP lines in the outflow is consistent with gas whose kinematics are dominated by bulk outflow motion with a roughly uniform level of turbulence across ionisation phases, producing little variation with IP.

The mean average velocity dispersion for the low-IP, disk-tracing lines is $78 \pm 11$ km/s. For the three intermediate-IP lines that trace both disk and outflow components, it is $108 \pm 12$ km/s, and for the high-IP, purely outflow-tracing lines, $119 \pm 13$ km/s.

While fitting single-Gaussian profiles provides a useful first-order characterisation of the gas kinematics, particularly for capturing large-scale rotational trends, it breaks down in inner regions where multiple, distinct kinematic components contribute to the line emission, resulting in non-Gaussian profiles. Examination of the line shapes reveals that, for the three intermediate-IP lines, the majority of spaxels exhibit double-peaked profiles, suggesting emission arising from two kinematically decoupled components.

\subsection{Double-Gaussian fitting}
\label{sec:doublegauss}

To more thoroughly explore the kinematics traced by each emission line, we attempt to fit double-Gaussian profiles to each continuum-subtracted spectrum for every emission line in Table~\ref{tab:ion_lines}. Each fit comprises a sum of two Gaussians, each with the usual three free parameters (amplitude, centroid, and width), applied to the same spectral regions used in the single-Gaussian fitting. This approach yields two velocity maps per emission line, corresponding to the two fitted centroids.

\subsubsection{Low and high-IP lines: double-Gaussian}

We find that for most of the emission lines listed in Table~\ref{tab:ion_lines}, double-Gaussian fits typically assign one component to the dominant disk and outflow velocity field, while the second merely captures residual noise, producing unphysical velocity structures. This indicates that the majority of lines are well described by a single-Gaussian across most spaxels, particularly those tracing only the disk (Fig.~\ref{fig:ArII_spaxel}). In contrast, several high-IP outflow-tracing lines, most notably [S IV], [O IV], and [Ne V], are significantly better reproduced by a double-Gaussian (Fig.~\ref{fig:SIV_spaxel}), but with both components tracing the outflow kinematics, through having the same PA as that of the outflow (e.g., Fig.~\ref{fig:SIV_2maps}). While one might alternatively attribute one of these components to emission from the AGN narrow-line region, we find this unlikely, as both Gaussian components exhibit highly similar and coherent velocity fields aligned with the outflow axis. Furthermore, both components show similar spatial distributions across the FoV. A similar behaviour was reported by \citet{2022ApJ...925..203J}, who, leveraging a larger FoV, showed that both components are associated with the extended ionisation cone (due to limb brightening). Given the lack of discernible differences in the velocity structure and amplitude between the two components in our data, we also find no evidence that either component traces the narrow-line region.

A similar effect was reported by \citet{2022ApJ...925..203J}, who showed that in optical emission lines (H$\alpha$, H$\beta$, [N~II]~6548, 6584\AA, [S~II]~6717, 6731\AA, [O~III]~4959, 5007\AA), spectra extracted from a purely disk region were adequately fit with single-Gaussians, whereas spectra from outflow regions were moderately improved by double-Gaussian fits. Our observations align with these findings for the high-IP lines mentioned above. For other high-IP transitions, double-Gaussian fits do not produce two physical velocity fields; we find that one Gaussian component traces the outflow, and the other residual noise, but this effect may simply reflect their lower S/N compared to [O~IV], [S~IV], and [Ne~V]. Overall, these results align with the suggestion from \citet{2022ApJ...925..203J} that the double-Gaussian profiles in these high-IP lines likely arise because the ionisation cone is at least partially hollow. Hence, double-peaked Gaussian profiles would emerge naturally when both the near and far sides of the cone edge (one blueshifted, the other redshifted) are simultaneously visible along the line of sight of a given spaxel.

\subsubsection{Intermediate-IP lines: double-Gaussian}

Beyond the purely disk- and cone-tracing lines, the intermediate-IP transitions ([Ne~III], [Ar~III], [S~III]) are particularly notable. These lines are not only well reproduced by double-Gaussian fits, but each Gaussian component traces a distinct kinematic structure. Specifically, the disk rotation is generally associated with the narrower Gaussian, while the broader component corresponds to the outflow (Fig.~\ref{fig:NeIII_spaxel}), consistent with the trends seen in Fig.~\ref{fig:mean_and_median_dispersions} from the single-Gaussian analysis. On this basis, we defined the outflow component in each spaxel as the Gaussian with the larger velocity dispersion, while the disk is represented by the narrower component and reran the double-Gaussian fitting process. For clarity, we refer to these as component~1 (disk) and component~2 (cone), requiring $\sigma_{v-\text{disk}} < \sigma_{v-\text{cone}}$ for each spaxel.

\begin{figure}
   \centering
   \includegraphics[width=\columnwidth]{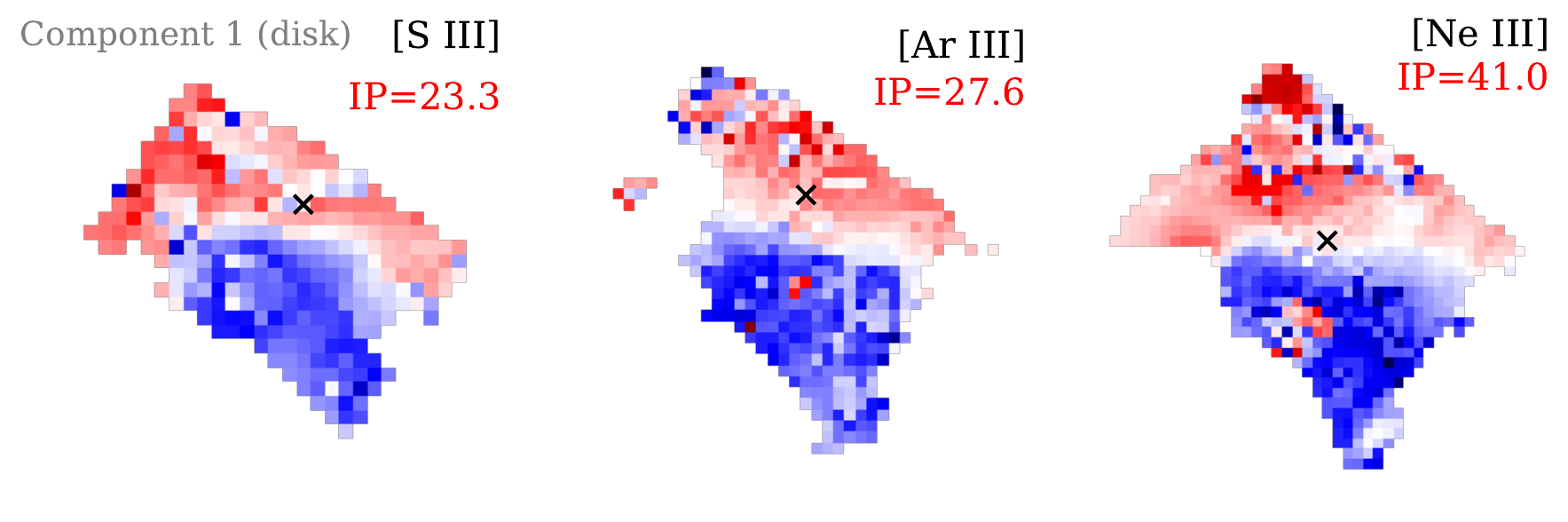}
   \includegraphics[width=\columnwidth]{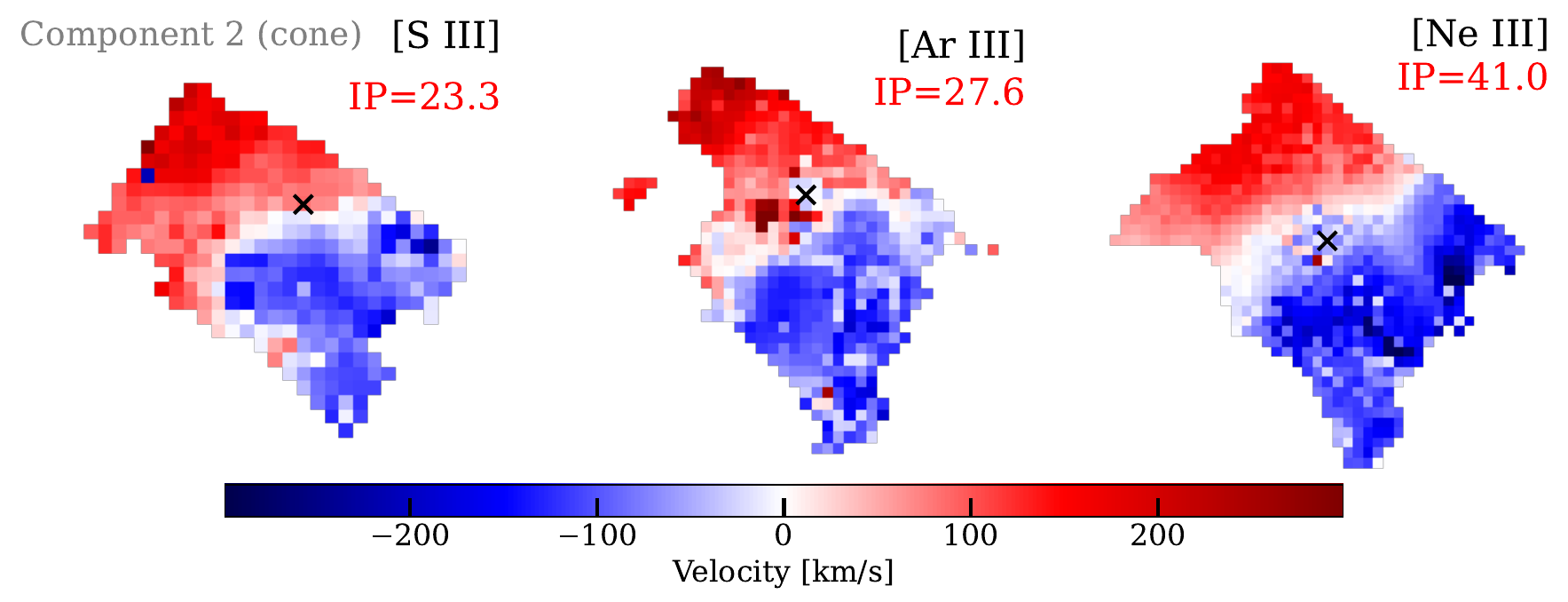}
   \caption{Velocity maps for the intermediate-IP emission lines in NGC~7582 fit using double-Gaussians ordered by IP (in eV). Top: disk Gaussian component (average major axis PA $\sim -13^\circ$), Bottom: cone (outflow) Gaussian component (average major axis PA $\sim 41^\circ$). The kinematics trace (disk or cone) was set based on the criteria $\sigma_{v-\text{disk}} < \sigma_{v-\text{cone}}$ for each spaxel. Crosses mark the AGN position. North is up, east is to the left.}
    \label{fig:double_gauss_vel_comp}%
\end{figure}

This classification produces two velocity maps for each line (Fig.~\ref{fig:double_gauss_vel_comp}) that reproduce the expected kinematic signatures: the disk component aligns with the major axis rotation seen in the low-IP lines, while the outflow component follows the outflow kinematics traced by the high-IP lines (Fig.~\ref{fig:single_gauss_vel_comp}). We show this quantitatively in Fig.~\ref{fig:PA_vs_IP} (right), where we show the position angles from PAFit for each Gaussian component of the intermediate-IP lines separately, with the disk and outflow component PAs aligning with what is seen from the solely disk/outflow tracing emission lines. Overall, these results provide strong evidence that the intermediate-IP lines have contributions from both the disk and outflow velocity fields.

\begin{figure*}
   \centering
   \includegraphics[width=0.33\linewidth]{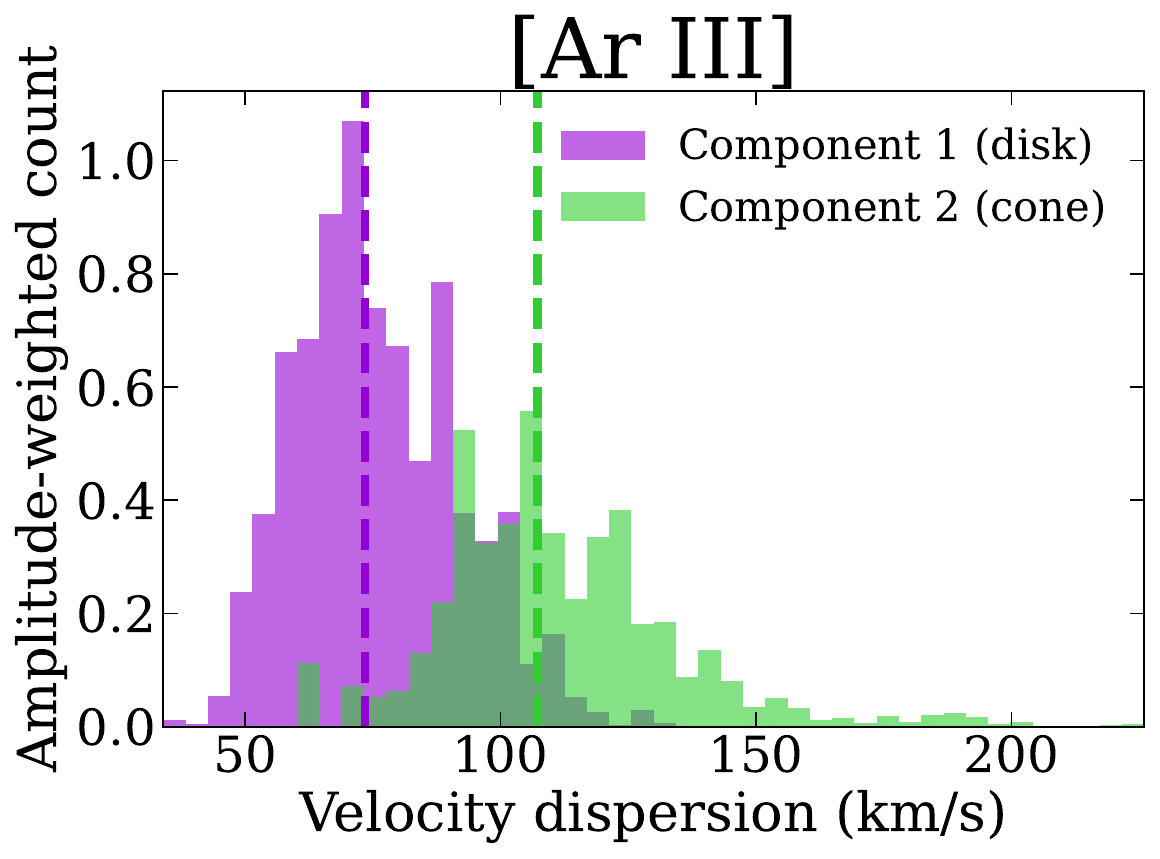}
   \includegraphics[width=0.33\linewidth]{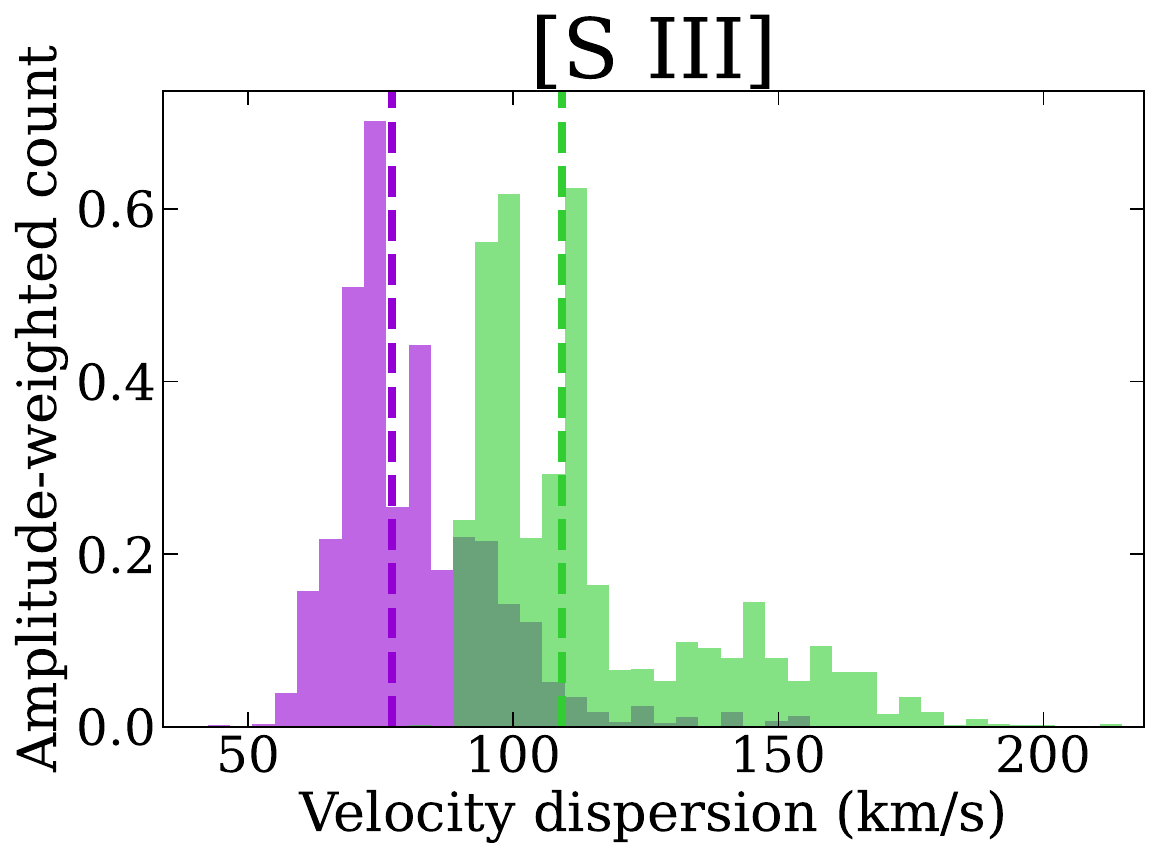}
   \includegraphics[width=0.33\linewidth]{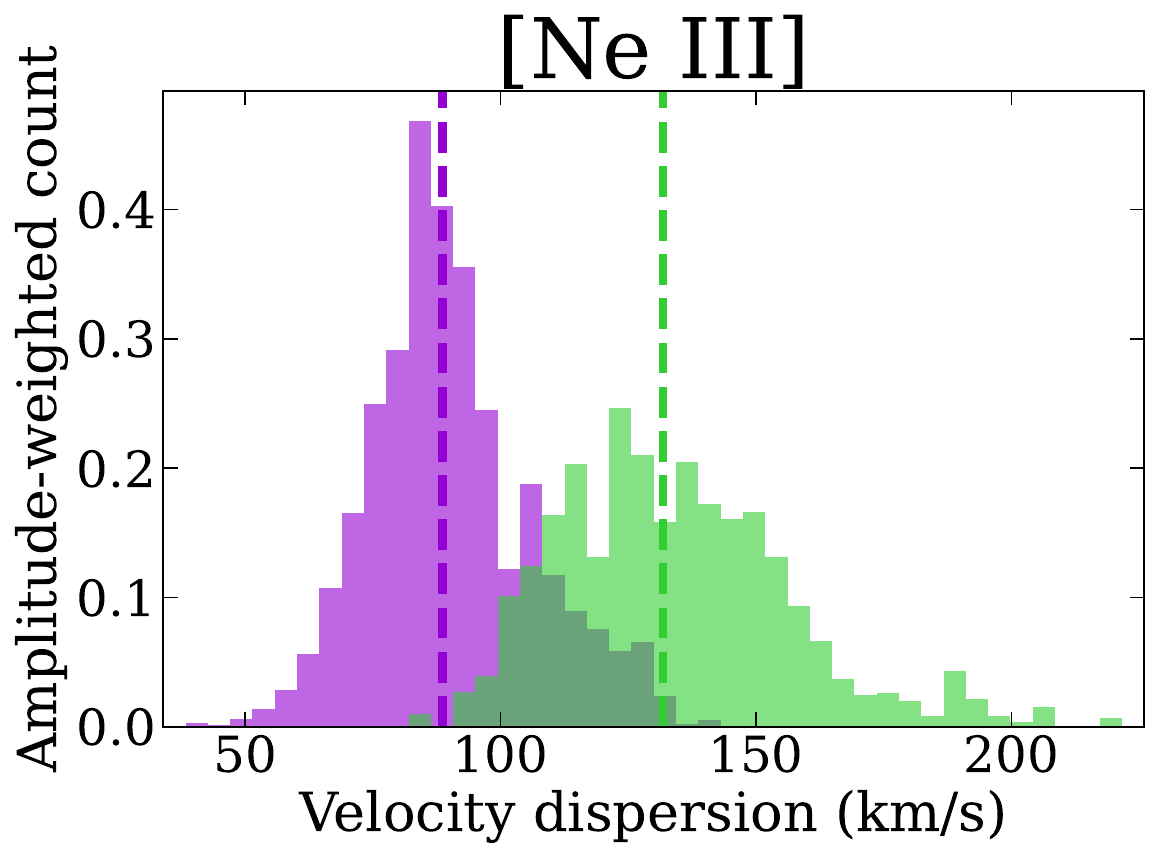}
   \caption{Gaussian amplitude weighted histograms of the distribution of the velocity dispersion of each Gaussian component in each spaxel for the intermediate-IP lines from double-Gaussian fitting. The disk dispersion distribution is shown in purple and the outflow dispersion distribution is shown in green. Dashed lines show the median of each histogram.}
    \label{fig:double_gauss_dispersion_histograms}%
\end{figure*}

We also examine the velocity dispersion distributions of each Gaussian component from the double-Gaussian fits, weighted by amplitude, for all three intermediate-IP lines (Fig.~\ref{fig:double_gauss_dispersion_histograms}). The histograms reveal two distinct populations (shown in purple and green for each component), with the median disk component dispersion consistently lower than that of the outflow component. Importantly, these two distributions remain distinct even without enforcing $\sigma_{v-\text{disk}} < \sigma_{v-\text{cone}}$, when components are instead classified by visual inspection of whether they trace the disk or the outflow, showing this separation is robust, and not simply due to construction. The disk component distributions are generally narrower with less scatter, indicative of ordered rotational kinematics. In contrast, the outflow components exhibit broader and more variable dispersions across the field of view, reflecting the turbulent nature of the ionised outflow.

We further refine our IP versus median velocity dispersion analysis for the intermediate-IP lines by plotting their median dispersions of each Gaussian component separately. As shown in Fig.~\ref{fig:mean_and_median_dispersions} (right), each intermediate-IP line now includes its median disk component (diamond) and average outflow component (square) velocity dispersion. The low-IP lines, along with the disk components of the intermediate-IP lines, occupy the lower-dispersion region of the plot, while the high-IP lines and outflow components lie at higher dispersions. Although separating the components introduces additional scatter, particularly among the low-IP group, this approach reinforces that gas associated with the circumnuclear disk exhibits systematically lower velocity dispersion than gas in the outflow. Consequently, the intermediate-IP lines continue to follow the same overall trends when their components are considered independently, strengthening the evidence that these transitions are excited by both the disk and outflow.

Further comparing the two Gaussian components across all spaxels for the three intermediate-IP lines, we find that the disk component amplitude is on average $\sim 50\%$ larger than that of the outflow component. However, this does not imply that most of the flux originates in the disk, since the total flux depends on the integrated line profile and the outflow components are generally broader as discussed (Fig.~\ref{fig:double_gauss_dispersion_histograms}). A similar result was reported by \citet{garcia2021multiphase}, who found that the outflow-tracing emission lines in NGC~5643 exhibit lower amplitudes and higher velocity dispersions than the emission lines associated with the disk. Overall, we measure comparable flux contributions (same order of magnitude) from both the disk and outflow components in the intermediate-IP lines. This balance is physically reasonable: if one component dominated the flux, the kinematic signature of the other would be much harder to detect.

\begin{figure}
   \centering
   \includegraphics[width=\columnwidth]{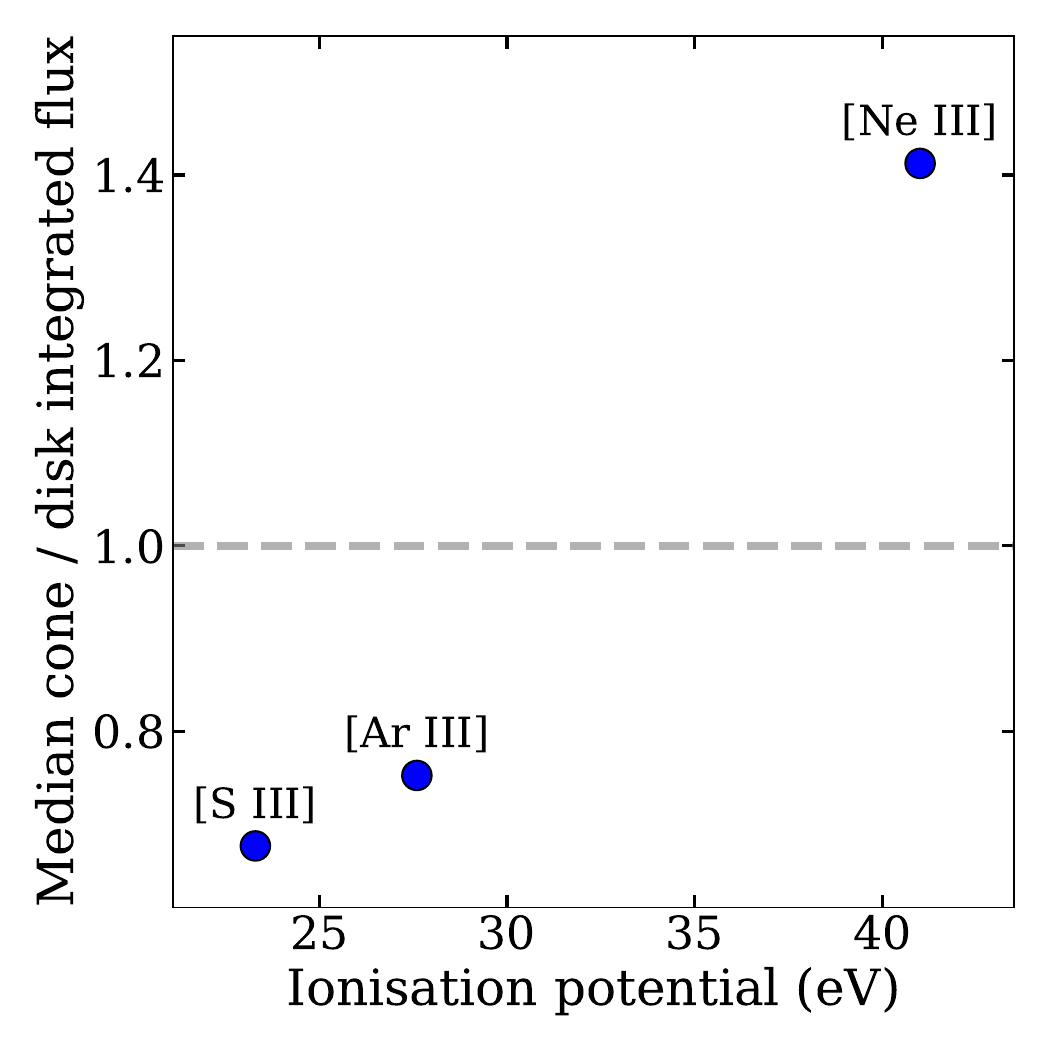}
   \caption{Median cone / disk integrated flux ratio across all unmasked spaxels for the intermediate-IP lines, showing that [S~III] and [Ar~III] are primarily excited in the disk, while [Ne~III] is mostly excited in the cone. However, cone and disk fluxes are comparable for all three lines, differing by only up to $\sim 50\%$ on average, indicating that each line can trace both components since neither dominates.}
    \label{fig:cone_disk_flux_ratios}%
\end{figure}

Fig.~\ref{fig:cone_disk_flux_ratios} presents the ratio of the median cone-to-disk Gaussian component flux, defined as $(\langle A_{\text{cone}}\rangle \langle \sigma_{\text{cone}}\rangle) / (\langle A_{\text{disk}}\rangle \langle \sigma_{\text{disk}}\rangle)$ (where $A$ and $\sigma$ are the Gaussian amplitude and dispersions respectively), across all unmasked spaxels for the intermediate-IP lines. This demonstrates that the fluxes of the cone and disk components are generally comparable, while revealing a dichotomy in disk versus cone properties with IP: [Ne III] shows more cone emission, whereas [S III] and [Ar III] are predominantly excited in the disk. These trends support our interpretation that the intermediate-IP lines are excited by both disk and outflow components, with the relative contributions strongly influenced by IP, however neither component strongly outshining the other for these lines.

As illustrated in Figs.~\ref{fig:single_gauss_vel_comp} and \ref{fig:PA_vs_IP}, the disk exhibits a kinematic major axis with PA~$\sim -12^\circ$, and previous studies \citep{riffel2009agn} have suggested that it is highly inclined and aligned close to perpendicular with the plane of the sky. This orientation results in a high column density, contributing to the significant dust obscuration of the AGN. X-ray studies report line-of-sight column densities of $N_{\mathrm{H}} \sim 10^{23} - 10^{24}$ cm$^{-2}$ for NGC~7582 \citep{rivers2015nustar}, consistent with a heavily obscured nucleus, while measurements also indicate substantial nuclear obscuration ($\tau_{9.8} \sim 2.67$; \citealt{veenema2025shock}).

In contrast, investigations into the geometry of the ionisation cone \citep{morris1985velocity, 2022ApJ...925..203J} indicate that it is inclined at a substantial angle relative to our line of sight, meaning we are not observing gas motion directly along the axis of maximum outflow velocity. This orientation difference leads to a projection effect: while the disk's velocity field is almost fully projected along our line of sight (thus yielding radial velocities that closely reflect the true maximum rotational speed), the cone's projected radial velocities underestimate the true outflow speeds. Consequently, although the outflow components of the intermediate-IP lines already show larger speeds than the disk components (as shown in Fig.~\ref{fig:single_gauss_vel_comp}), these values are likely lower limits due to its inclination.

The double-Gaussian fitting has been highly informative, confirming that most emission lines primarily trace a single dynamical component, with only the three intermediate-IP lines showing strong signatures of both disk rotation and outflow motion. However, in some spaxels the method reaches its limits, making it difficult to fully separate and map the kinematics of both components. For instance, in the [Ne III] disk component velocity map shown in Fig.~\ref{fig:double_gauss_vel_comp}, several spaxels located to the north and south of the zero velocity axis deviate significantly from the expected disk rotation pattern and from the behaviour of surrounding spaxels. We note that these apparent deviations in the [Ne~III] disk component are likely not indicative of true departures from ordered rotation, but rather reflect limitations of the moderately unconstrained double-Gaussian fitting in spaxels where the two components overlap. This highlights the need for more physically motivated modelling to robustly separate disk and outflow kinematics. Similar noisy regions are also seen in the [Ar III] and [S III] component velocity maps. Once more, these discrepancies likely also reflect spaxels where the fitting routine has struggled to accurately recover both components due to overlapping Gaussians that result in an overall line profile that resembles close to a single-Gaussian. This issue is exacerbated by the nature of automated fitting procedures, particularly when using models with a relatively high number of free parameters (six, in the case of the double-Gaussian fit), which can introduce degeneracies and lead to unreliable solutions in some spaxels. Thus, although the double-Gaussian fitting has performed surprisingly well overall in separating disk and outflow kinematics in the intermediate-IP lines, a more physically motivated and constrained modelling approach is warranted. To this end, we proceed by fitting an inclined thin disk rotation model to the data and subsequently fitting the then isolated cone component, and vice versa using a simplified, first-order cone model. This strategy reduces the number of free parameters in each case, and may allow for more robust, decoupled, and physically meaningful kinematic decomposition.

\subsection{Disk modelling}
\label{sec:diskmodelling}

\subsubsection{The thin inclined rotating disk}

In this subsection, we aim to model the kinematics of the disk component independently, then subtract its contribution from the total intermediate-IP emission line profiles. This approach allows us to isolate and analyse the residual line profiles under the assumption that they primarily trace the kinematics of the outflowing gas. To proceed, we first define and fit a physically motivated model for the rotation of the circumnuclear disk. We adopt a thin, inclined rotating disk model, centred on the AGN position, with free parameters such as the inclination of the disk relative to our line of sight, $i$, and the PA of its major kinematic axis. These parameters must be either constrained from the data or fixed based on independent observational estimates to enable a meaningful fit. To model the disk, we analyse several low-IP emission lines (Fig.~\ref{fig:single_gauss_vel_comp}).

We previously verified the PA of the disk's major axis using PAFit \citep{Krajnovic2006} on the single-Gaussian velocity maps, finding a PA$\sim -12 \pm 3^\circ$ being highly consistent across all low-IP lines (Fig.~\ref{fig:PA_vs_IP}).

\begin{figure*}
   \centering
   \includegraphics[width=\linewidth]{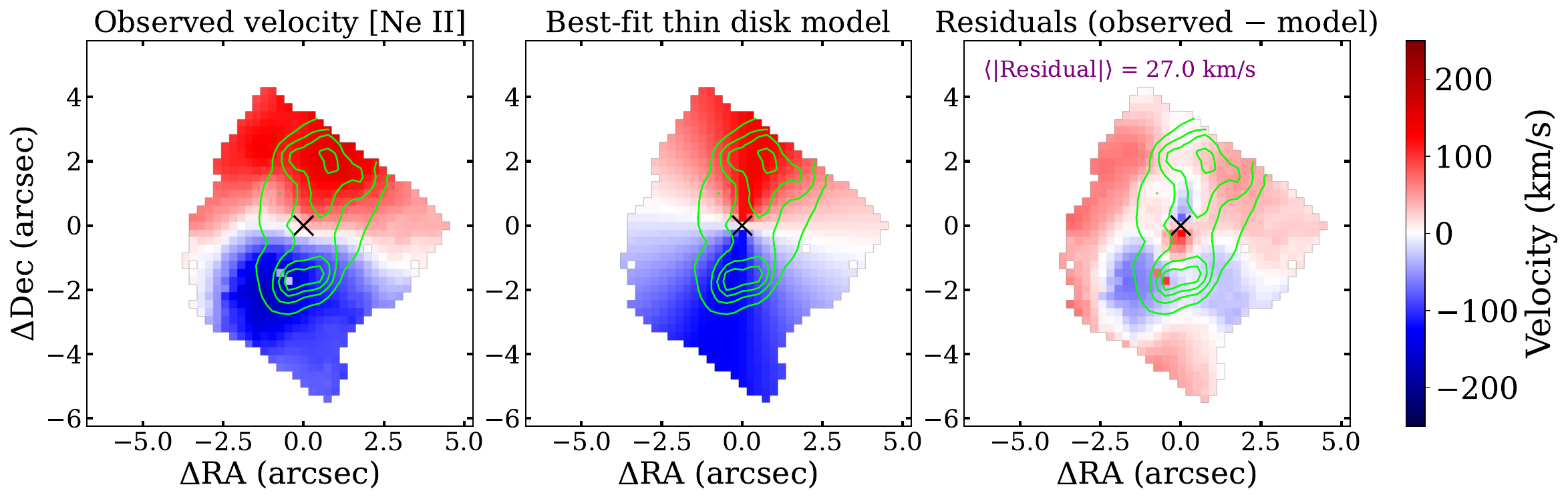}
   \includegraphics[width=\linewidth]{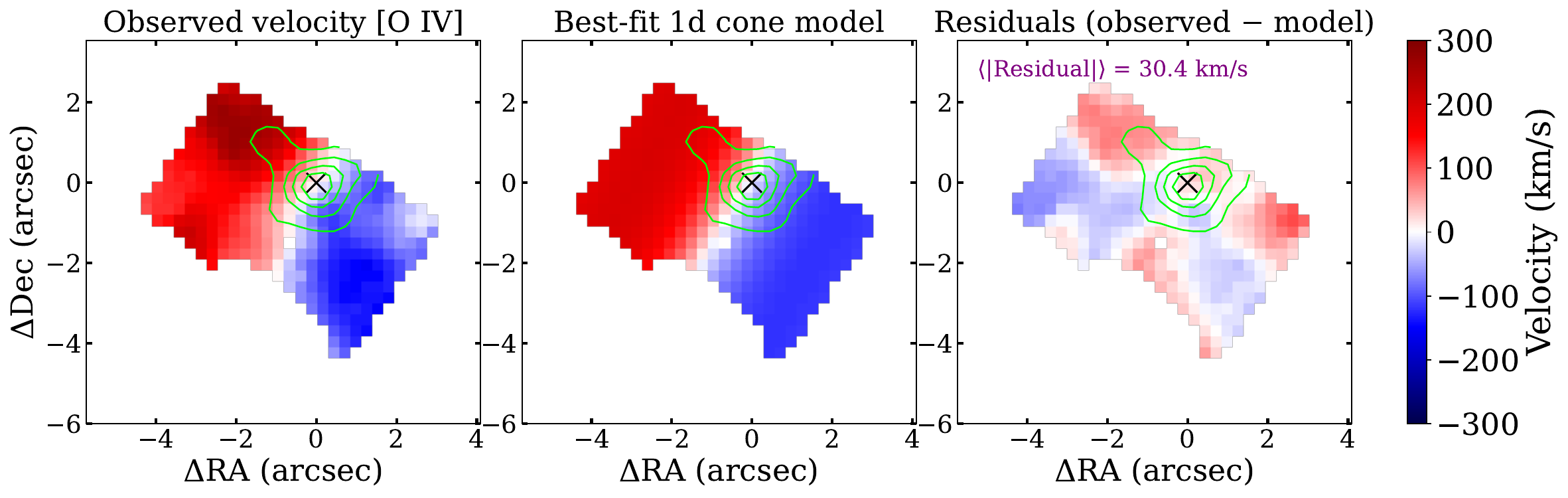}
   \caption{Top left: [Ne II] single-Gaussian velocity map, tracing the circumnuclear disk rotation. Top middle: Inclined thin rotating disk model velocity field from the [Ne II] velocity map with best-fit parameters inclination, $i=58^{\circ}$, major velocity axis PA $= -12^{\circ}$, and maximum rotational velocity ($V_{\mathrm{max}}$) and turnover radius ($R_{\mathrm{turn}}$) as given in Table~\ref{tab:vmax_rturn}. Top right: Residuals of the model taken away from the observations (top left - top middle) with the mean absolute residual given, which is lower than the instrumental velocity dispersion in this MIRI/MRS channel. Green contours on the top plots show the [Ne II] integrated flux emission, which strongly traces the star-forming disk. Bottom left: [O IV] single-Gaussian velocity map. Bottom middle: Best-fit outflow model from [O IV]. Bottom right: Residual (bottom left - bottom middle) map with the mean absolute residual shown, which is lower than the instrumental resolution in this MIRI/MRS channel. The green contours show the [O IV] integrated flux emission which is mostly PSF dominated from the AGN emission, with some elongation along the ionisation cone axis direction. The cross denotes the AGN position.}
    \label{fig:NeII_disk_velocity_map}%
\end{figure*}

For the disk inclination relative to the line of sight, $i$, we fix this parameter at $i = 58^\circ$, consistent with values derived by previous studies \citep{morris1985velocity, wold2006nuclear2, garcia2021galaxy}. While more sophisticated fitting techniques such as 3DBAROLO \citep{teodoro20153d} could be employed to independently constrain the inclination, we find that adopting this fixed value yields a disk rotation model that reproduces the observed velocity fields exceptionally well (see Fig.~\ref{fig:NeII_disk_velocity_map} (top) and its discussion), and we have seen no major discrepancies between any of our velocity analyses that suggest large differences with the literature.

With the PA and inclination fixed, we proceed to fit the projected two-dimensional rotational velocity field for a thin disk, where the projected sky coordinates $(x, y)$ are measured relative to the galaxy centre. We adopt a tanh rotation curve to describe the line of sight velocity field, which captures the typical shape of a disk rotation profile rising steeply near the centre and flattening at larger radii, and gives a good empirical fit to our data. The functional form of the observed disk velocity field is therefore given by:

\begin{equation}
v_{\text{model}}(x, y) = v_{\text{sys}} + V_{\text{max}} \tanh\left(\frac{R}{R_{\text{turn}}}\right) \sin(i) \cos(\theta) .
\end{equation}

Here, $v_{\text{sys}}$ is the systemic velocity of the galaxy due to the Hubble expansion. For nearby galaxies, this can be approximated as $v_{\text{sys}} \approx cz$, where $c$ is the speed of light and $z$ is the redshift. For NGC~7582, we adopt $z = 0.00525$ \citep{braito2017high}. However, we note that all kinematic maps presented in this work have already been corrected by subtracting this systemic velocity. In the model, $R$ denotes the deprojected radius in the disk plane, and $\theta$ is the corresponding azimuthal angle in the disk plane, defined as:

\begin{align}
R &= \sqrt{x'^2 + \left( \frac{y'}{\cos i} \right)^2} \\
\theta &= \arctan\left( \frac{y'}{\cos i \cdot x'} \right) \\
x' &= x \cos\phi + y \sin\phi \\
y' &= -x \sin\phi + y \cos\phi .
\end{align}

where $\phi$ is the PA of the kinematic major axis (fixed at -12$^\circ$), and $(x', y')$ are the rotated $(x, y)$ coordinates aligned with the major axis.

Two additional free parameters in the model are the maximum (asymptotic) rotational velocity, $V_{\text{max}}$, and the turnover radius, $R_{\text{turn}}$, which sets the radial scale over the velocity profile transitions. These parameters are determined through an LM optimisation procedure, applied to the low-IP single-Gaussian velocity maps under the assumption that they purely trace the disk rotation, with the PA and $i$ fixed as described previously. To check consistency, we also performed fits using this model allowing the PA and $i$ to vary, and found that they converged to values consistent with our adopted PA = $-12^\circ$ and $i = 58^\circ$ within one standard error. We therefore fixed these parameters in all subsequent model fits to avoid potential degeneracies, given our confidence in their values. The resulting projected velocity field of the model disk from fitting to the [Ne II] velocity map is shown in Fig.~\ref{fig:NeII_disk_velocity_map} (top). We ran this procedure for all disk-tracing emission lines, with the best fit $V_{\text{max}}$ and $R_{\text{turn}}$ for each stated in Table~\ref{tab:vmax_rturn}. The values of $R_{\mathrm{turn}}$ and $V_{\mathrm{max}}$ are broadly consistent across all disk-tracing lines confirming that they are all tracing the same kinematics. We choose the [Ne II] parameters for our model moving forward as it has the highest S/N of all disk-tracing lines. Fitting this model to the [Ne II] single-Gaussian velocity map yields a reduced chi-squared of $\chi_r^2 = 2.22$.

\begin{table}
    \centering
    \caption[]{Best-fit rotation curve parameters; maximum rotational velocity ($V_{\mathrm{max}}$) and the turnover radius ($R_{\mathrm{turn}}$), derived from thin disk modelling for each disk-tracing emission line. Uncertainties are the 3$\sigma$ errors.}
    \label{tab:vmax_rturn}
    \begin{tabular}{lcc}
        \hline
        Emission Line & $V_{\mathrm{max}}$ (km/s) & $R_{\mathrm{turn}}$ (pc) \\
        \hline
        {[Fe II]} & 178 $\pm$ 9 & 86 $\pm$ 10 \\
        {[Cl II]} & 177 $\pm$ 9 & 128 $\pm$ 20 \\
        HI 7-5 & 147 $\pm$ 9 & 75 $\pm$ 10 \\
        {[Ar II]} & 146 $\pm$ 9 & 107 $\pm$ 10 \\
        {[Ne II]} & 166 $\pm$ 12 & 118 $\pm$ 20 \\
        \hline
    \end{tabular}
\end{table}

\subsubsection{Disk model application and results}

With an analytical velocity field for the circumnuclear disk established, Fig.~\ref{fig:NeII_disk_velocity_map} (top), we can incorporate it into the analysis of the intermediate-IP line kinematics. In its most straightforward application, the model disk velocity field provides a prediction for the centroid wavelength of the disk component in an emission line profile. If the double-Gaussian line profile assumption is retained, this effectively fixes one of the model parameters - namely the disk centroid, $\lambda_{\text{disk}}$, reducing the number of free parameters to five: the centroid, dispersion, and amplitude of the outflow component, $\lambda_{\text{outflow}}, \sigma_{\text{outflow}}, A_{\text{outflow}}$, and the dispersion and amplitude of the disk component, $\sigma_{\text{disk}}, A_{\text{disk}}$.

While the approach of fixing only the centroid wavelength of the disk component provides a basic method for aiding in isolating the outflow contribution in intermediate-IP lines, it may not yield the most accurate results as there are still five free parameters. An alternative strategy involves also constraining one or both of the remaining disk Gaussian parameters too, informed by insights from our earlier double-Gaussian fits. As previously discussed, we found that the disk component typically exhibited an emission line amplitude $\sim 50\%$ higher than that of the outflow component on average. Furthermore, as shown in Fig.~\ref{fig:double_gauss_dispersion_histograms}, the disk velocity dispersion is systematically lower than that of the outflow and approximately follows a Gaussian distribution across all spaxels.

These findings suggest it may be justifiable to fix either or both of $\sigma_{\text{disk}}$ and $A_{\text{disk}}$ to representative values. For instance, $\sigma_{\text{disk}}$ could be set to the median dispersion from the disk component of the double-Gaussian fits (i.e. the dotted purple lines in Fig.~\ref{fig:double_gauss_dispersion_histograms}), and $A_{\text{disk}}$ to 1.5$A_{\text{outflow}}$. Although fixing these parameters might marginally reduce the fidelity of the disk component model, it mitigates degeneracies commonly encountered in automated fitting routines. This, in turn, facilitates more reliable convergence on the outflow kinematics by minimising contamination from the disk component in the intermediate-IP line profiles.

For completeness, and given the difficulty in intuitively determining which approach best decouples the kinematics, we test all methods and present the resulting velocity fields for the three intermediate-IP lines in Fig.~\ref{fig:cone_fits_from_disk_model}, where each subfigure shows the resulting outflow velocity field found from fitting a single-Gaussian to it after initially subtracting the disk model applied in different ways. Each row shows the results for a specific intermediate-IP line, and each column shows a different model fitting approach as given by its title. These disk models all adopt the AGN position as the kinematic centre and share the same parameters as the thin rotating disk model previously derived from the [Ne II] velocity field but of course projected onto the corresponding NIRSpec or MIRI/MRS FoV and spaxel scale for each specific line.

Overall, the derived outflow velocity fields for each intermediate-IP line (Fig.~\ref{fig:cone_fits_from_disk_model}) exhibit the expected morphology, with PAs consistent within 1$\sigma$ errors with the high-IP, outflow tracing velocity maps from the single-Gaussian fits (PA~$\sim 57^\circ$). When compared to the outflow fields obtained from the full six parameter double-Gaussian fits (bottom row of Fig.~\ref{fig:double_gauss_vel_comp}), the results from most fixed disk parameter combinations are less noisy. This improvement is particularly evident in the case of [Ne III], which provides the clearest example of successfully decoupled disk and outflow kinematics.

The most reliable outflow fits across each intermediate-IP line (Fig.~\ref{fig:cone_fits_from_disk_model}) are obtained either by fixing all disk-Gaussian parameters (1st column) or by allowing only the disk amplitude to vary (3rd column). These approaches yield velocity maps with minimal noise and far fewer spaxels that diverge drastically from the kinematics of their surroundings. This demonstrates that fixing the disk velocity dispersion to the median value for each line, as determined from the initial double-Gaussian fits (Fig.~\ref{fig:double_gauss_dispersion_histograms}), is crucial for robustly recovering the outflow velocities. In contrast, when $\sigma_{v-\text{disk}}$ is left free (2nd and 4th columns in Fig.~\ref{fig:cone_fits_from_disk_model}), we observe a significant increase in noisy or poorly constrained spaxels, especially in [Ar III] and [S III]. We therefore conclude that $\sigma_{v-\text{disk}}$ introduces substantial degeneracies, and fixing it to a physically motivated value is both appropriate and necessary to cleanly separate disk rotation from outflow kinematics when applying a thin inclined disk model.

This result is consistent with \citet{bellocchi2019uncertainties}, who showed that the velocity dispersion is the parameter most sensitive to fitting methodology, while the velocity centroid remains comparatively robust. Allowing $\sigma$ to vary introduces larger systematic deviations and increased scatter, particularly at low S/N. Their results support our choice to fix $\sigma_{v-\mathrm{disk}}$ to a physically motivated value in order to minimise degeneracies and obtain more reliable gas kinematics.

\subsection{Biconical outflow modelling}
\label{sec:conemodelling}

We have shown that incorporating a fixed disk model to constrain two or three parameters provides a more coherent picture of both the disk and outflow velocity fields in the intermediate-IP lines than a fully unconstrained double-Gaussian fit. An alternative strategy is to reverse the fitting order by first modelling the outflow kinematics independently, then fitting the disk Gaussian profile to the residuals. We explore this approach in this subsection.

\subsubsection{1st order outflow model}

Unlike ordered disk rotation, modelling outflow kinematics is less straightforward, particularly on the sub-kiloparsec scales probed closest to the AGN in this study. Ionised outflows, including the one in NGC~7582, are thought to be driven by AGN accretion powered radiation, which becomes collimated by the dusty torus, producing axially symmetric outflows perpendicular to the torus plane \citep{garcia2021galaxy, 2022ApJ...925..203J}. Gas in the ISM that lies along the path of these outflows becomes strongly ionised as it absorbs this radiative energy from the AGN. Observationally, it is unclear whether more powerful outflows are more frequently detected in heavily dust-obscured Type 2 AGN, \citep{tozzi2024super, 2025MNRAS.542.2525H}, nonetheless suggesting that interactions between accreted dust and the torus play a key role in launching them, which has been shown also through simulations \citep{soliman2023dust}. This makes modelling outflows challenging: understanding of their physical origin is incomplete, and their kinematic signatures are less intuitive than those of rotating disks.

We begin by analysing the kinematics of the outflow-tracing high-IP lines. Many such lines are present in the NIRSpec and MIRI/MRS wavelength ranges, but the cleanest tracer of the outflow with the best S/N is [O IV], as shown in our single-Gaussian velocity maps in Fig.~\ref{fig:single_gauss_vel_comp}. On these scales, the kinematics in the [O IV] map might, at a glance, resemble disk-like rotation, with a major kinematic PA of roughly $42^\circ$. The maps in Fig.~\ref{fig:single_gauss_vel_comp} are flux-masked, showing only spaxels above the 60th percentile in flux for each line, thus the morphology of each subplot reflects the spatial extent of the emission. For [O~IV] and other high-IP transitions, the structure appears to originate at the AGN and broadens with distance, as seen from the increasing number of unmasked spaxels further along the kinematic major axis. The emission aligns with the known ionisation cone rather than the disk major axis, and the velocity dispersions are systematically larger than in disk-tracing lines, consistent with turbulent or bulk motions in an outflow. Moreover, the high-IP ions require hard radiation fields and so should be preferentially produced in the AGN photoionised cone, not from stellar feedback in the circumnuclear disk. Finally, the rotation-like appearance arises from the limited FoV ($\sim$200~pc), whereas the cones extend to $\gtrsim$3~kpc \citep{2022ApJ...925..203J, marconcini2023moka3d}. Together, these arguments show that the high-IP lines do not trace a disk rotation, but instead the inner regions of the AGN-driven ionised outflow.

\begin{figure}
   \centering
   \includegraphics[width=\columnwidth]{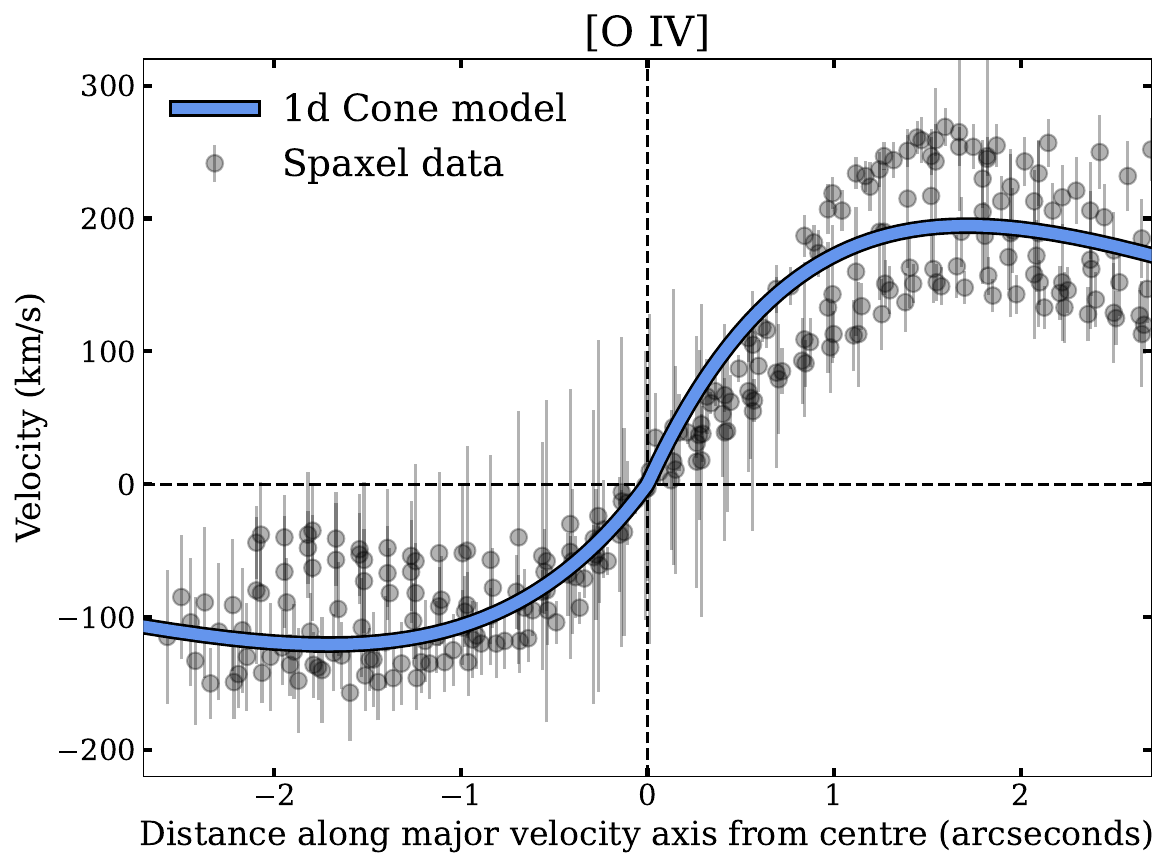}
   \caption{[O IV] single-Gaussian fit velocity for each spaxel as a function of distance along the outflow kinematic major axis (PA$\sim 42^\circ$ from PAFit) from the AGN. The blue line shows the best fitting asymmetrical exponential decay model which we use to then construct our analytical 1d outflow velocity field shown in Fig.~\ref{fig:NeII_disk_velocity_map} bottom.}
    \label{fig:OIV_SIV_velocity_curve_fits}%
\end{figure}

We present in Fig.~\ref{fig:OIV_SIV_velocity_curve_fits} the velocity of each spaxel, derived from single-Gaussian fits, as a function of distance from the AGN along the kinematic major axis for [O IV]. This profile reveals a strong dependence of gas velocity on position along the major axis of the ionisation cone. The gas reaches a peak velocity on either side of the nucleus. This peak is particularly pronounced and occurs between approximately 1-2$\arcsec$ from the kinematic centre, beyond which the speed gradually declines. We also see asymmetry between the redshifted and blueshifted sides of the cone, with the redshifted side exhibiting maximum speeds that are $\sim 50$ km/s greater. This might indicate stronger acceleration on the far side of the cone or interaction with the galactic disk. We also examined the velocity structure as a function of distance perpendicular to the cone axis, i.e. toward the cone edges, and found no statistically significant variation across the field of view. This suggests a relatively uniform acceleration mechanism throughout the ionisation cone, with no enhanced mechanical energy injection at larger angular distances from the cone axis. For consistency, we also checked this for [S IV] and [Ne V] (not shown), and confirm we see the same profile shape and trends as discussed for [O IV].

Several analytical formulae for AGN outflow kinematics have been adopted (e.g., \citealt{das2005mapping, das2007dynamics, mullersanchez2011, fischer2013determining, durre2019agn}), typically featuring a monotonic rise in velocity over a few hundred parsecs to a maximum speed, followed by a decline toward the galaxy's systemic velocity by $\sim1$ kpc. The high-IP lines in NGC~7582 show a similar pattern: the outflow accelerates to a peak and then gradually decelerates. However, the maximum speeds on the redshifted and blueshifted sides differ, necessitating an asymmetrical profile to more accurately capture the outflow velocity field heuristically.

Overall, as a simple approach, we model the outflow velocity field as a simple one-dimensional (1d) velocity field, where velocities vary as a function of distance along the major axis of the cone from the AGN. This can be determined by fitting an asymmetrical decaying profile to the velocity curve shown in Fig.~\ref{fig:OIV_SIV_velocity_curve_fits}. Mathematically, this is expressed as:

\begin{equation}
v(r) =
\begin{cases}
v_{\text{red}} \cdot \left( \dfrac{r}{r_d} \right) \exp\left( -\dfrac{r}{2 r_d} \right), & r \geq 0 \\
v_{\text{blue}} \cdot \left( \dfrac{r}{r_d} \right) \exp\left( \dfrac{r}{2 r_d} \right), & r < 0
\end{cases}
\label{eq:asym_exp_disk}
\end{equation}

where $r$ denotes the distance along the major kinematic axis of the cone, and $v_{\text{red}}$, $v_{\text{blue}}$, and $r_d$ represent the maximum asymptotic speeds on the redshifted and blueshifted sides, and the scale over which the velocity profile rises (the same for both the blue and red shifted sides), respectively. We find that allowing for asymmetric maximum speeds on either side of the cone significantly improves the fit, with equation~(\ref{eq:asym_exp_disk}) outperforming asymmetric tanh and arctan profiles. Since our goal is to empirically characterise the velocity field of the outflow, this fitting formula is both sufficient and appropriate for our purposes.

For our model based on the [O IV] kinematics, we obtain a best-fit of $v_{\text{red}} = 260 \pm30$ km/s, $v_{\text{blue}} = -160\pm30$ km/s, and $r_d = 130\pm20$ pc. Fitting this model to the [O IV] single-Gaussian velocity map yields a reduced chi-squared of $\chi_r^2 = 3.39$ (which is worse than the disk model to the low-IP lines, and likely due to our outflow model being more simplistic, with less free parameters). When applied to the [Ne V] and [S IV] velocity profiles, we find consistency within the two-$\sigma$ uncertainties. Therefore, we adopt these parameters for all subsequent outflow modelling.

\subsubsection{Outflow model application and results}

Our 1d outflow model can then be directly compared to the observed [O IV] velocity map, as shown in Fig.~\ref{fig:NeII_disk_velocity_map} (bottom), and then reprojected onto the IFS FoV of other line observations to account for the outflow velocity component as desired. We find that the residuals between this simple 1d model and the [O IV] single-Gaussian velocity map used to generate it are small, typically comparable to or below the level of instrumental velocity dispersion, and that the model reproduces the observed velocities well as shown by Fig.~\ref{fig:NeII_disk_velocity_map} bottom.

We can create model outflow velocity maps for the intermediate-IP lines and, analogously to the previous section, use this to keep various outflow Gaussian parameters fixed. We use the median velocity dispersion from the outflow Gaussian distribution (green dashed lines in Fig.~\ref{fig:double_gauss_dispersion_histograms}), in fits where the outflow dispersion is fixed, and for fixed amplitude, we again set $A_{\text{disk}}$ to 1.5$A_{\text{outflow}}$, finding this typically gives the best fits. The resulting disk velocity fits for the different parameter combinations and for the different intermediate-IP lines are presented in Fig.~\ref{fig:disk_fits_from_cone_model}.

These maps demonstrate that subtracting the outflow component in different ways prior to fitting the disk Gaussian yields velocity maps that closely trace the expected disk kinematics. For all three intermediate-IP lines, this procedure produces velocity maps with the same overall morphology as those derived from the disk Gaussian component of the full double-Gaussian fits (Fig.~\ref{fig:double_gauss_vel_comp}), but with much reduced noise. This highlights that this approach is also more robust than a full six-parameter double-Gaussian fit. We ran PAFit on the recovered disk velocity field and find a position angle consistent with a PA$\sim -12^\circ$ within two standard errors for all three lines in all four methods presented in Fig.~\ref{fig:disk_fits_from_cone_model}, confirming that we are accurately resolving the disk. We have therefore demonstrated that, by first modelling the outflow in a simple 1d prescription, the intrinsic disk velocity field can be reliably recovered.

We observe the same trend as in the disk modelling used to fit the outflow (Fig.~\ref{fig:cone_fits_from_disk_model}). Specifically, the most successful fits in Fig.~\ref{fig:disk_fits_from_cone_model} occur in the 1st and 3rd columns, where either all outflow model parameters are fixed or only the outflow amplitude is allowed to vary. In contrast, permitting the outflow velocity dispersion to vary consistently produces poorer overall fits to the recovered disk velocity maps for each intermediate-IP line. Thus, in both approaches - whether subtracting the disk to fit the outflow, or subtracting the outflow to fit the disk - our results indicate that allowing both velocity dispersions to vary simultaneously significantly degrades the fits. It is therefore advantageous to fix at least one velocity dispersion parameter to a representative value in order to achieve more reliable kinematic decompositions. This should be an appropriate physically representative value, such as the median of the velocity dispersion distributions found from the double-Gaussian fits (Fig.~\ref{fig:double_gauss_dispersion_histograms}).

\subsection{Simultaneous disk and outflow constraints}
\label{sec:hybridmodel}

Our disk-first and outflow-first modelling approaches show that the parameters exerting the strongest influence on the derived velocity fields are the velocity dispersions of the disk and outflow, $\sigma_{v-\text{disk}}$ and $\sigma_{v-\text{outflow}}$ respectively. To reduce the degeneracy and constrain the modelling, we now fix these dispersions and fit only the remaining four parameters of the double-Gaussian model: the amplitudes and centroid wavelengths of the outflow and disk components. We adopt, for each intermediate-IP line, the median dispersion values obtained from the fully unconstrained double-Gaussian fits (Fig.~\ref{fig:double_gauss_dispersion_histograms}). With these dispersions fixed, we re-fit the data allowing the other four parameters to vary freely to derive the corresponding disk and outflow velocity maps, shown in Fig.~\ref{fig:hybrid_double_gauss_vel_comp}.

These revised velocity maps from double-Gaussian fitting represent a notable improvement over the fully unconstrained fits shown in Fig.~\ref{fig:double_gauss_vel_comp} for all intermediate-IP lines. In particular, the derived disk velocity fields exhibit substantially reduced noise and more easily identifiable position angles. The outflow kinematics also show higher velocities relative to the disk.

The outflow spaxels exhibit higher noise than those in the disk due to fixing the velocity dispersion. As shown in Fig.~\ref{fig:double_gauss_dispersion_histograms}, the outflow has a broader velocity dispersion distribution, with fewer spaxels near the median compared to the disk. Consequently, assuming a constant dispersion is less accurate for the outflow and introduces noise across more spaxels than it does for the disk.

\subsection{Escaping gas within the outflow?}
\label{sec:escape_velocity}

In this section we further examine the properties of the gas in the ionised outflow. \citet{wold2006nuclear2} modelled the gravitational potential of the circumnuclear stellar bulge (i.e. the star-forming disk) of NGC~7582 using a multi-Gaussian expansion \citep{2002MNRAS.333..400C} fit to the NICMOS F160W image. From their Table~1, the summed luminosity of the four bulge components is $2.06\times10^{9} L_\odot$. Adopting their stellar mass-to-light ratio of 3.8 gives a total stellar bulge mass of $7.83\times10^{9} M_\odot$. They also estimated the central SMBH mass to be $\sim 5.5\times10^{7} M_\odot$, yielding a total enclosed mass within the inner~$\sim$200 pc of $7.89\times10^{9} M_\odot$. This estimate neglects the contributions from gas and dark matter, which are assumed to be negligible compared to the stellar component (disk + bulge) and SMBH masses on these spatial scales. Using this mass, we estimate the escape velocity at a distance of 200 pc from the AGN to be $v_{\text{esc}} \sim 580$ km/s.

From our outflow velocity modelling (Fig.~\ref{fig:OIV_SIV_velocity_curve_fits}), the ionised gas within the cone reaches projected velocities of up to $\sim200$ km/s at this radius. Because the outflow is a hollow bicone with a finite opening angle, individual gas parcels follow different trajectories and therefore intersect our line of sight at slightly different inclinations. As a result, the observed velocity field represents a superposition of projected velocities spanning a range of true outflow directions, rather than a single inclination. This naturally produces the observed spread in projected velocities, yielding average values of $\sim 200$ km/s at 200~pc for the mid-IR outflow-tracing lines. The corresponding deprojected velocities along the cone axis are higher, but remain highly sensitive to the inclination of the specific outflowing gas relative to our line of sight.

Several studies have estimated the outflow inclination in NGC~7582. Notably, \citet{marconcini2025fast} performed a detailed kinematic analysis of ionised AGN winds using optical emission lines, finding a best-fit inclination of $i \sim 70^\circ$ for the blueshifted outflow. Similarly, \citet{lopez2020amusing}, modelling nearby galaxies with VLT/MUSE IFS data, found $i \sim 72^\circ$. These estimates are in good agreement and broadly consistent with the ionisation cone models of \citet{morris1985velocity} and \citet{ricci2018optical}, which, although not quantitatively constraining the inclination, support a similarly inclined geometry. Adopting $i \sim 70-72^\circ$ implies that the deprojected outflow velocities are roughly a factor of 3 higher than the observed values. The mid-IR outflow-tracing lines analysed here exhibit typical velocities of $\sim 200$ km/s at 200~pc, corresponding to true deprojected velocities of $\sim 600$ km/s along the cone axis.

This indicates that a significant fraction of gas may escape the gravitational potential of the galaxy and be deposited into the circumgalactic (CGM) or intergalactic medium (IGM) rather than being re-accreted. Indeed, \citet{marconcini2025fast} examined the ratio of outflow velocity to escape velocity as a function of distance from the AGN using tracers at visible wavelengths (their fig.~3), finding values of $\sim 1$-$1.5$, in good agreement with the deprojected mid-IR outflow velocity to escape velocity ratio inferred here.

\subsection{The effect of IP on the opening angle of the AGN ionisation cone in context of the dusty torus}
\label{sec:opening_angle_vs_IP}

We further investigate the morphology of the AGN ionisation cones and their connection to the properties of the unresolved dusty torus by analysing how the cone opening angle may be affected by IP for each high-IP line. The study by \citet{2022ApJ...925..203J} reported a hollow ionisation cone structure, for which we found supporting evidence in the form of double-peaked emission line profiles in several high-IP lines, with both Gaussian components tracing the outflow velocity field.

We now examine the hollow cone in greater detail by analysing the high-IP line flux maps to identify signatures of the cone edges as a function of angle at a fixed distance from the AGN. The edges of the hollow cone appear brighter than its interior, consistent with a structure in which the bicone is largely evacuated because gas inside is being rapidly driven outwards or compressed toward the cone walls by the outflow. This morphology is visible in many high-IP flux maps shown in Fig.~\ref{fig:flux_comp}.

\begin{figure}
   \centering
   \includegraphics[width=\columnwidth]{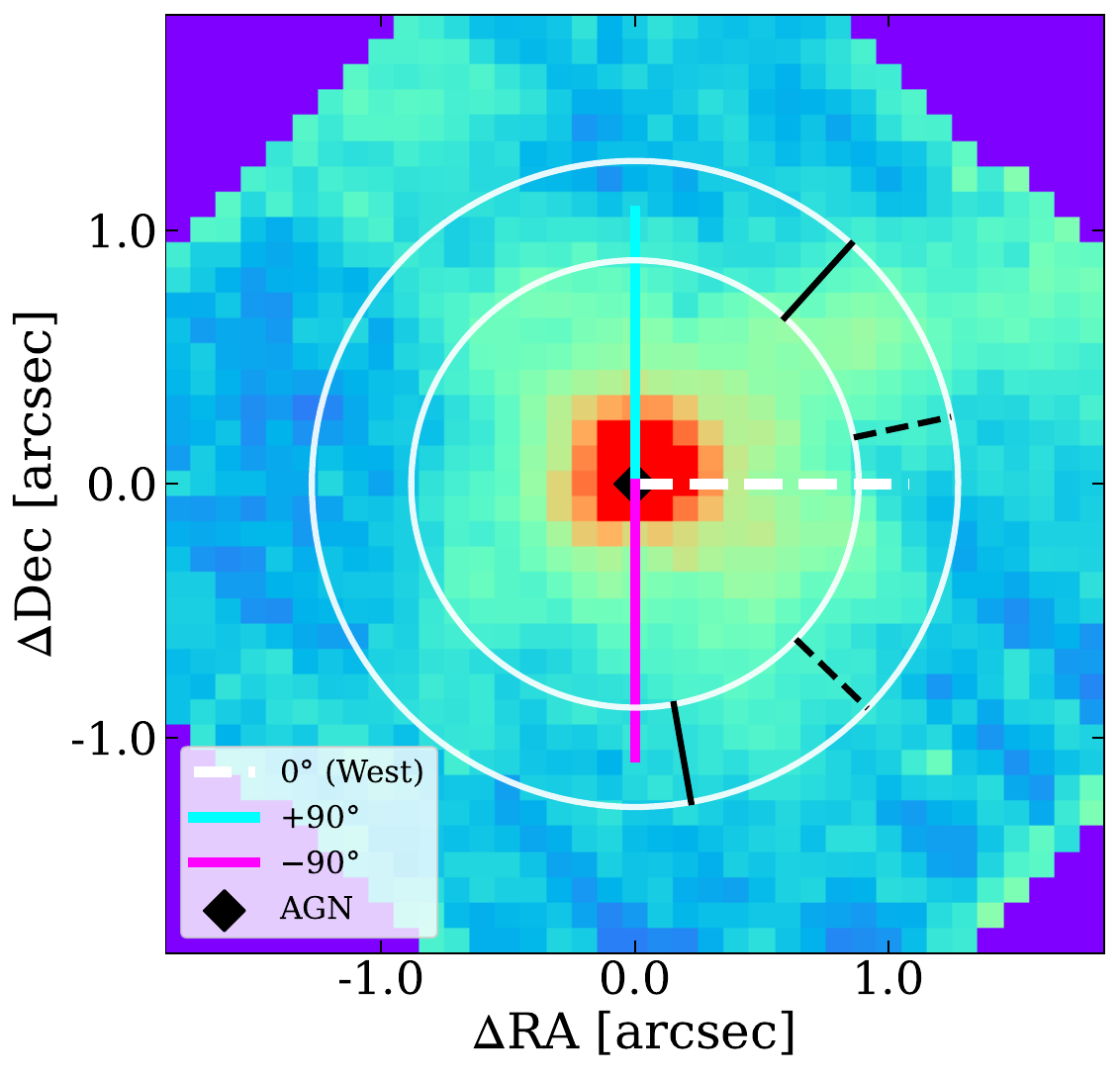}
   \includegraphics[width=\columnwidth]{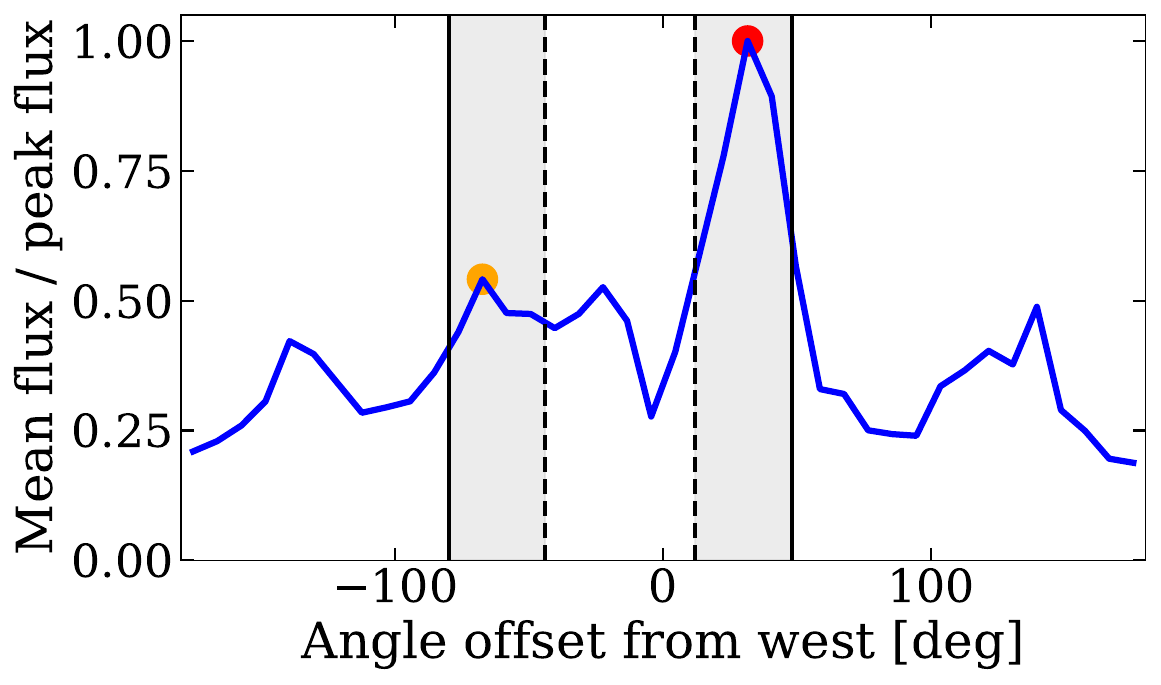}
   \caption{Top: [Mg IV] flux map showing the annulus considered and directionality of the angles. The colour scaling has been cut to 3 times below the maximum value to highlight the edges of the ionisation cone. Bottom: average flux as a function of angle throughout the annulus for the [Mg IV] flux with the cone edge regions shaded and the positive peak identified as the red circle, and the negative peak identified as the orange circle. The dashed and solid black lines enclose the angular regions where we expect the cone edge to lie on both plots, with the dashed showing the inside of the edge and solid showing the outer extent of the edge. The two unmarked peaks beyond $\pm 100^\circ$ correspond to the edges of the eastern cone which is obscured by the disk.}
    \label{fig:MgIV_cones}%
\end{figure}

To investigate this further, we analyse the flux within a hypothetical arc of fixed width at a constant distance from the AGN (i.e. an annulus). We adopt an inner radius of 0.8$\arcsec$ ($\sim 80$\:pc) and an outer radius of 1.2$\arcsec$ ($\sim 120$\:pc). This choice provides a suitable compromise to be sufficiently distant from the AGN to minimise PSF contamination, even in the longest MIRI/MRS channel, while also remaining fully within the FoV of all emission lines at all position angles. The annulus is illustrated in Fig.~\ref{fig:MgIV_cones} (top) using the [Mg IV] integrated flux map as an example, with its boundaries shown with white solid lines, and the $\pm$90$^\circ$ directions marked by cyan and purple lines, respectively. For convenience, we define west as the zero-angle reference, though this choice does not affect the results that follow.

The edges of the ionisation cones were identified as bright and kinematically distinct features in the optical [O III] line by \citet{2022ApJ...925..203J} (their figures 4, 6, and 7). Based on their results, we estimate the angular positions of the upper (northern) and lower (southern) edges of the westward, front-facing ionisation cone. To account for uncertainty, we include a $\pm$5$^\circ$ margin around each edge. The resulting regions correspond to +12$^\circ$ to +48$^\circ$ and -44$^\circ$ to -80$^\circ$ from west for the northern and southern edges, respectively. These are shown graphically in Fig.~\ref{fig:MgIV_cones} (top) as black lines, where solid lines denote the outer boundaries and dashed lines the inner boundaries of each edge.

We then examine the average flux as a function of angle within the annulus. As an example, Fig.~\ref{fig:MgIV_cones} (bottom) shows this distribution for [Mg IV], where the flux is normalised to the peak average flux within the annulus at each angle. The same edge positions are indicated by the black lines, and the shaded grey regions between each pair of lines mark the angular ranges where the cone-edge emission is expected to peak. The angles corresponding to the maximum mean flux within each region are identified as the cone-edge positions, marked by the red circle for the upper edge and the orange circle for the lower edge.

This analysis is repeated for all high-IP lines using the same annulus (fixed dimensions for inner and outer radius from the AGN), identifying the upper and lower cone edges as the flux peaks within the grey regions. For brevity, we show only the [Mg IV] result in Fig.~\ref{fig:MgIV_cones}, as all high-IP lines exhibit similar flux-angle profiles for this annulus. In all cases, the upper cone edge displays the strongest peak, while the lower edge is consistently weaker.

\begin{figure}
   \centering
   \includegraphics[width=\columnwidth]{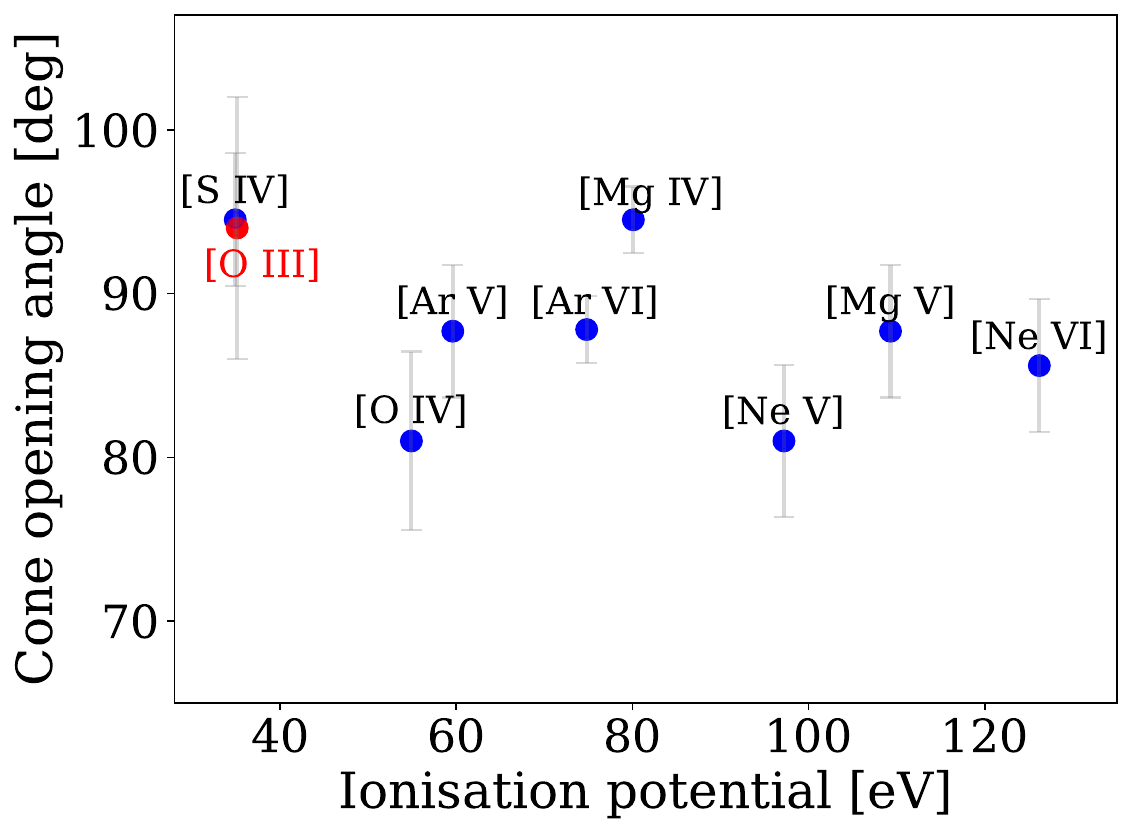}
   \caption{Cone opening angle (angle between positive and negative peaks) for each high-IP line plotted against their IP, including the [O III] 5007 line estimated from \citet{2022ApJ...925..203J}. We see no clear trend, indicating the cone opening angle is not affected by IP, i.e. that the AGN torus is thick to all emission lines and thus is not stratified in obscuring material density with IP.}
    \label{fig:cone_opening_angle_vs_IP}%
\end{figure}

We calculate the cone opening angle as the angular separation between the two edge peaks, equal to the difference in angles of the red and orange circles for each high-IP line. The resulting opening angles are plotted as a function of IP in Fig.~\ref{fig:cone_opening_angle_vs_IP}, with the uncertainties derived from the finite angular resolution set by the spaxel size of each IFU. We also include the [O III] 5007 line (IP = 35.1 eV) estimated from its flux map presented by \citet{2022ApJ...925..203J} (their fig. 4) as an additional point in red.

Fig.~\ref{fig:cone_opening_angle_vs_IP} shows no clear trend between IP and the measured cone opening angle, which remains consistently between $\sim80$-$95^\circ$, with average opening angle across all nine emission lines being $88\pm3^\circ$. Although a linear fit to the data points yields a slight negative gradient, this trend is not statistically significant within the associated uncertainties. We therefore conclude that the biconical outflow maintains a constant opening angle across all high-IP lines. This implies that the AGN torus is relatively geometrically well-defined, constraining the escape of ionising photons similarly across this energy range, consistent with a torus in which the polar regions are largely unobscured.

This result does not require the torus to have a smooth, steeply declining density profile, as a clumpy torus with optically thick clouds would produce a similar constant opening angle \citep{nenkova2008agni, nenkova2008agnii}. In such a scenario, photons escaping along the polar axis always encounter the same effective optical depth, regardless of the detailed cloud distribution.

\begin{figure}
   \centering
   \includegraphics[width=\columnwidth]{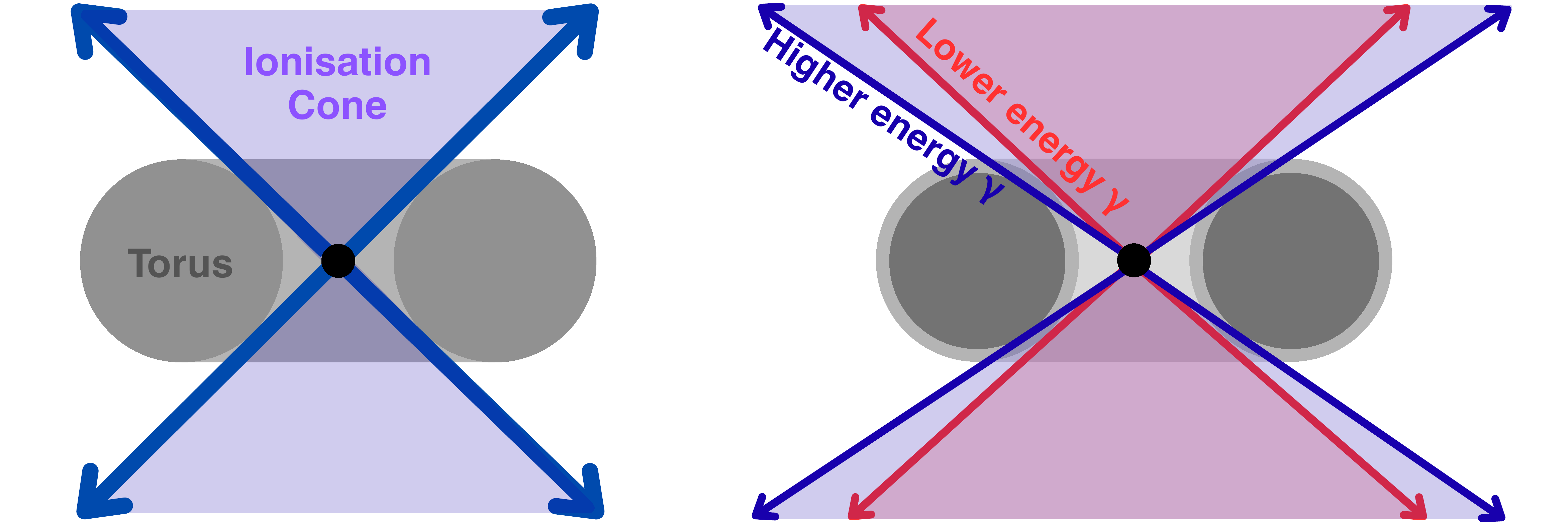}
   \caption{Schematic illustration of the two possible AGN torus morphologies considered. Each panel shows a cross-section of the torus with the central SMBH at the centre, and arrows indicating the boundaries of observable emission from the hollow ionisation cones. Left: A torus with a polar axis that is largely unobscured, allowing ionising photons of all energies to escape along similar angles, producing ionisation cones with a roughly constant opening angle across IP. Right: A torus in which the polar axis contains more obscuring material, such as a two-phase medium with a low-density component, which would preferentially block low-IP photons and produce a narrower cone for lower-energy emission lines. Note that a clumpy torus could produce the same effects, provided a low-density filling component is present in the second case.}
    \label{fig:Torus_schematic}%
\end{figure}

However, our results do constrain the torus central region: the polar axis must contain relatively little obscuring material, allowing photons of all energies to escape along similar angles. This corresponds to the torus configuration on the left in Fig.~\ref{fig:Torus_schematic}. A torus filled with a two-phase medium including a low-density component along the axis, as in the schematic on the right, would produce a narrower low-IP bicone, which is not observed.

\section{Conclusions}
\label{sec:conclusions}

In this paper, we analysed the inner regions of NGC~7582 using JWST NIRSpec and MIRI/MRS integral-field spectroscopic data, focusing on the kinematics of ionic emission lines based on their IP. Our goal was to disentangle and characterise the circumnuclear disk/ring rotation from the AGN-driven outflow, in the context of an active galaxy with complex nuclear morphology. Our key findings are as follows:

\begin{enumerate} 

\item The circumnuclear gas in NGC~7582 is kinematically stratified: low-IP emission lines trace the ordered circumnuclear disk/ring rotation with a PA$\sim -12 \pm 3^\circ$, while high-IP lines predominantly follow the AGN photoionised biconical outflow with a PA$\sim 54 \pm 10^\circ$ (Fig.~\ref{fig:PA_vs_IP}).

\item The outflow-tracing gas exhibits systematically higher velocity dispersions than the disk-tracing gas on average and even on a per spaxel basis, consistent with turbulent or bulk motions in an ionised wind compared to ordered rotation (Fig.~\ref{fig:double_gauss_dispersion_histograms}).

\item For lines tracing the disk component, velocity dispersion increases with IP, whereas the outflow shows a less well-defined trend, with a weak positive correlation between dispersion and IP (Fig.~\ref{fig:mean_and_median_dispersions}). 

\item The low-IP emission lines are well described by single-Gaussian profiles, whereas the high-IP lines with the highest S/N ([S IV], [O IV], and [Ne V]) show moderately improved fits when modelled with two Gaussian components. Both components trace the outflow kinematics, consistent with emission arising from a hollow ionisation cone.

\item Three intermediate-IP species ([Ne III], [Ar III], and [S III]) are better fit by a double-Gaussian profile, but exhibit contributions from both kinematic components (Fig.~\ref{fig:double_gauss_vel_comp}). The outflow component is characterised by $\sim 50\%$ higher velocity dispersion, $\sim 50\%$ lower amplitude, and higher maximum speeds on average compared to the disk component.

\item Kinematic models; either a thin inclined rotating disk or a simple 1d outflow model are remarkably effective at isolating the complementary kinematic component in the intermediate-IP lines, allowing us to decouple the disk from the outflow kinematics (Fig.~\ref{fig:NeII_disk_velocity_map}). We find that fixing the velocity dispersion of at least one of the components moderately improves the fit, leading to clearer derived velocity fields.

\item We compare the observed outflow velocities with the gravitational potential of the SMBH and stellar bulge at 200 pc (as calculated by \citealt{wold2006nuclear2}), finding that the gas reaches deprojected speeds comparable to the local escape velocity ($\sim600$ versus $\sim 580$ km/s). This suggests that the outflow is capable of ejecting material into the CGM or even the IGM, where it is unlikely to be re-accreted.

\item By examining the angular flux distribution at a fixed distance from the AGN for nine high-IP lines, we measure the opening angle of the front-facing ionisation cone and find no dependence on IP, with average opening angle of $88\pm3^\circ$. This constant opening angle with IP indicates that the polar regions of the AGN torus contain relatively little obscuring material, allowing ionising photons of all energies to escape along similar ranges in angle.

\end{enumerate}

Finally, this work has demonstrated the exceptional capabilities of JWST IFS for studying galactic kinematics in nearby galaxies in the mid-IR. The combination of reduced extinction at these wavelengths and JWST's sensitivity allows for the most accurate and precise measurements of gas kinematics to date.

\section*{Acknowledgements}

OV is supported by a Science and Technology Facilities Council (STFC) studentship No.\ ST/Y509474/1.
IGB is supported by the Programa Atracci\'on de Talento Investigador ``C\'esar Nombela'' via grant 2023-T1/TEC-29030 funded by the Community of Madrid, and acknowledges support from the research project PID2024-159902NA-I00 funded by the Spanish Ministry of Science and Innovation / State Agency of Research (MCIN/AEI/10.13039/501100011033) and FSE+.
SGB acknowledges support from the Spanish grant PID2022-138560NB-I00, funded by MCIN/AEI/10.13039/501100011033/FEDER, EU.
MPS acknowledges support under grants RYC2021-033094-I, CNS2023-145506, and PID2023-146667NB-I00 funded by MCIN/AEI/10.13039/501100011033 and the European Union NextGenerationEU/PRTR.
RAR acknowledges support from Conselho Nacional de Desenvolvimento Cient\'ifico e Tecnol\'ogico (CNPq; Proj.\ 303450/2022-3, 403398/2023-1, and 441722/2023-7), Funda\c c\~ao de Amparo \`a Pesquisa do Estado do Rio Grande do Sul (FAPERGS; Proj.\ 21/2551-0002018-0), and Coordena\c c\~ao de Aperfei\c coamento de Pessoal de N\'ivel Superior (CAPES; Proj.\ 88887.894973/2023-00).
AJB acknowledges funding from the ``FirstGalaxies'' Advanced Grant from the European Research Council (ERC) under the European Union's Horizon 2020 research and innovation programme (Grant agreement No.\ 789056).
SC and DJR acknowledge support from the UK's Science and Technology Facilities Council (STFC) through grant ST/X001105/1.
EB acknowledges support from the Spanish grants PID2022-138621NB-I00 and PID2021-123417OB-I00, funded by MCIN/AEI/10.13039/501100011033/FEDER, EU.
CRA and AA acknowledge support from the Agencia Estatal de Investigaci\'on of the Ministerio de Ciencia, Innovaci\'on y Universidades (MCIU/AEI) under the grant ``Tracking active galactic nuclei feedback from parsec to kiloparsec scales'', with reference PID2022--141105NB--I00 and the European Regional Development Fund (ERDF).
AA acknowledges support from the European Union (WIDERA ExGal-Twin, GA~101158446).
EKSH acknowledges grant support from the Space Telescope Science Institute (ID JWST-GO-03535).
SFH acknowledges support through UK Research and Innovation (UKRI) under the UK government's Horizon Europe funding guarantee (EP/Z533920/1) and an STFC Small Award (ST/Y001656/1).
AAH and LHM acknowledge support from grant PID2021-124665NB-I00 funded by the Spanish Ministry of Science and Innovation and the State Agency of Research MCIN/AEI/10.13039/501100011033 and ERDF (``A way of making Europe'').
OGM acknowledges financial support from the UNAM PAPIIT project IN109123 and CONAHCyT Ciencia de Frontera project CF-2023-G-100.

%%%%%%%%%%%%%%%%%%%%%%%%%%%%%%%%%%%%%%%%%%%%%%%%%%
\section*{Data Availability}

 The data cubes used here are publicly available from the Mikulski Archive for Space Telescopes (MAST)\footnote{\url{https://mast.stsci.edu/portal/Mashup/Clients/Mast/Portal.html}}, with DOI: 10.17909/9sc5-q436.



\bibliographystyle{mnras}
\bibliography{references} % if your bibtex file is called example.bib

%\begin{thebibliography}{99}
%\end{thebibliography}

%%%%%%%%%%%%%%%%%%%%%%%%%%%%%%%%%%%%%%%%%%%%%%%%%%

%%%%%%%%%%%%%%%%% APPENDICES %%%%%%%%%%%%%%%%%%%%%

\appendix

\section{Example of spectra and fits for individual spaxels}

\begin{figure}
   \centering
   \includegraphics[width=\columnwidth]{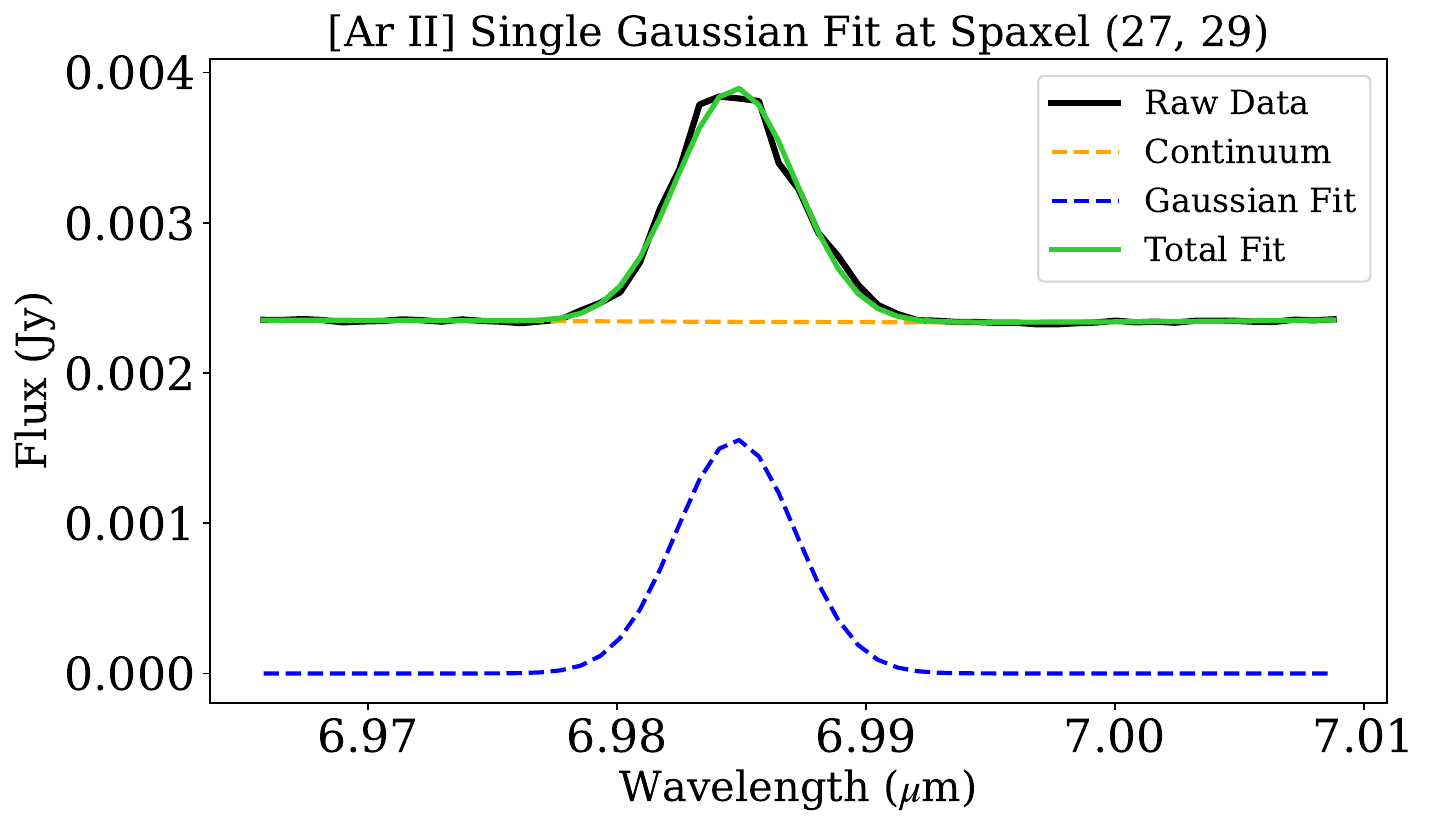}
   \caption{Representative single spaxel spectra around the [Ar II] line. This line strongly traces the circumnuclear star-forming disk and is well fit by a single-Gaussian plus continuum fit.}
    \label{fig:ArII_spaxel}%
\end{figure}

\begin{figure}
   \centering
   \includegraphics[width=\columnwidth]{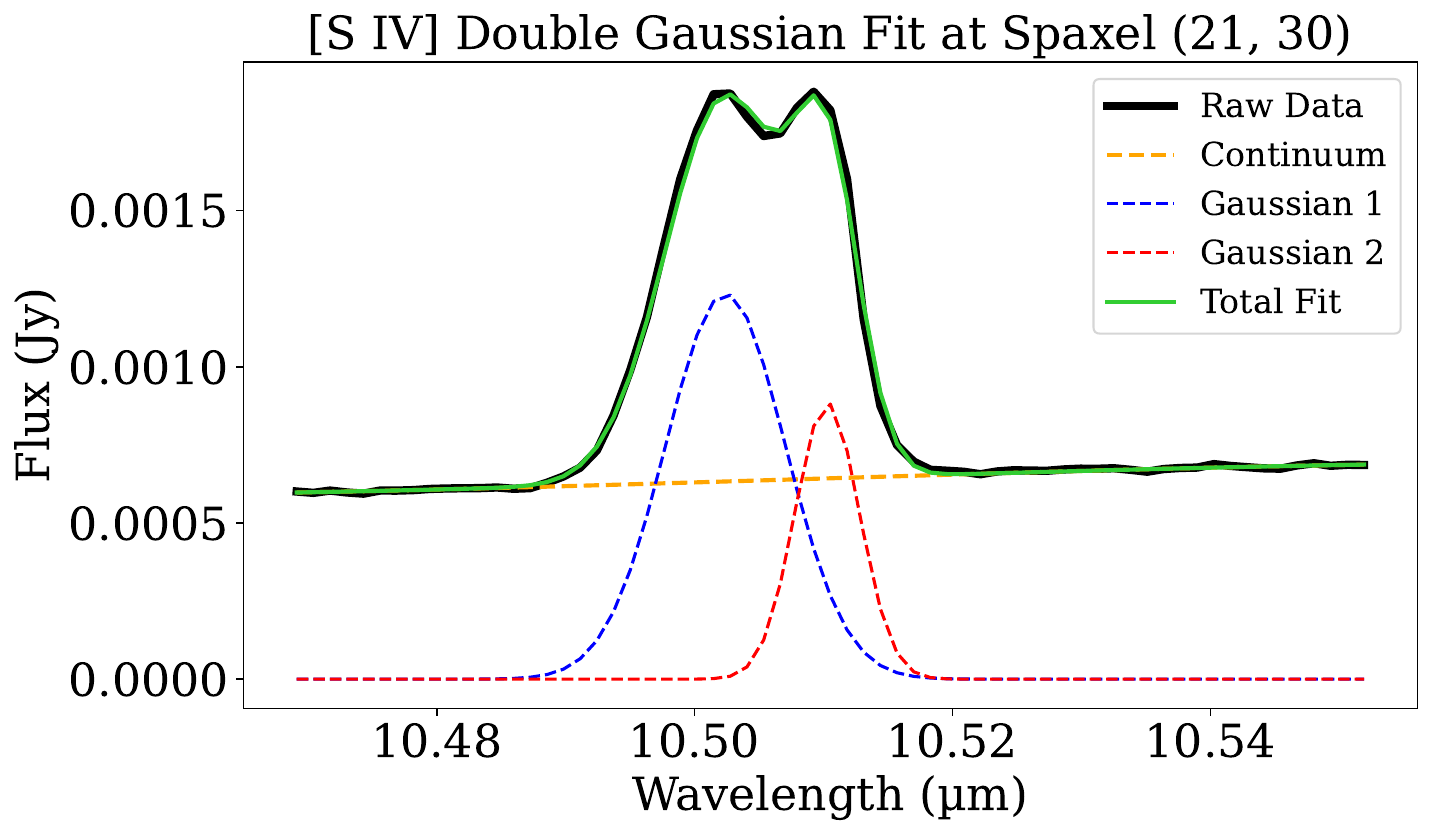}
   \caption{Representative single spaxel spectra around the [S IV] line. This line strongly traces the AGN driven outflow, and a double-Gaussian plus continuum profile provides a significantly better fit than a single-Gaussian plus continuum. Both best-fit Gaussian functions trace the outflow kinematics (see Fig.~\ref{fig:SIV_2maps}).}
    \label{fig:SIV_spaxel}%
\end{figure}

\begin{figure}
   \centering
   \includegraphics[width=\columnwidth]{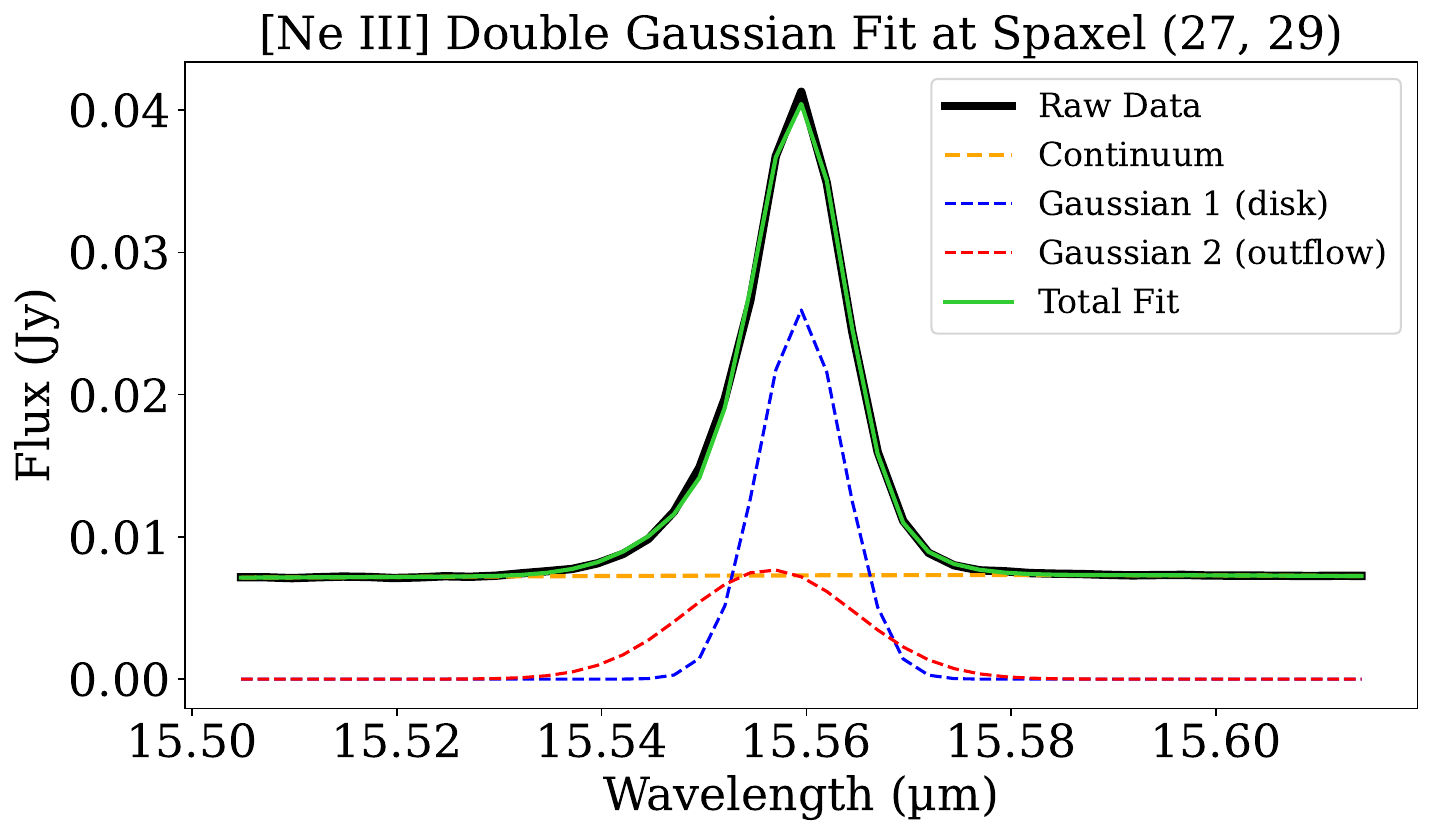}
   \caption{Representative single spaxel spectra around the [Ne III] line. This line strongly traces both the circumnuclear star-forming disk and AGN driven outflow, and a double-Gaussian plus continuum profile provides a significantly better fit than a single-Gaussian plus continuum. However, these two Gaussian fits differ in that one has a significantly higher amplitude, lower width (dispersion), and strongly follows the disk kinematics, whereas the other has significantly lower amplitude, larger width (dispersion), and follows the outflow kinematics.}
    \label{fig:NeIII_spaxel}%
\end{figure}

We present representative single-spaxel spectra and Gaussian fits illustrating the characteristic profiles of disk-tracing, outflow-tracing, and mixed disk/outflow-tracing emission lines. Specifically, we show examples for [Ar II], a primary disk tracer, [S IV], a strong outflow tracer, and [Ne III], an intermediate-IP line tracing both components (Figs.~\ref{fig:ArII_spaxel}, \ref{fig:SIV_spaxel}, and \ref{fig:NeIII_spaxel}, respectively).

Fig.~\ref{fig:ArII_spaxel} demonstrates that typical [Ar II] spaxel spectra (and other disk-tracing lines) are well described by a single-Gaussian profile plus continuum. Introducing a second Gaussian component does not significantly improve the fit, as the overall $\chi^{2}$ decreases by less than unity, indicating no statistically meaningful gain in goodness of fit.

\begin{figure}
   \centering
   \includegraphics[width=\columnwidth]{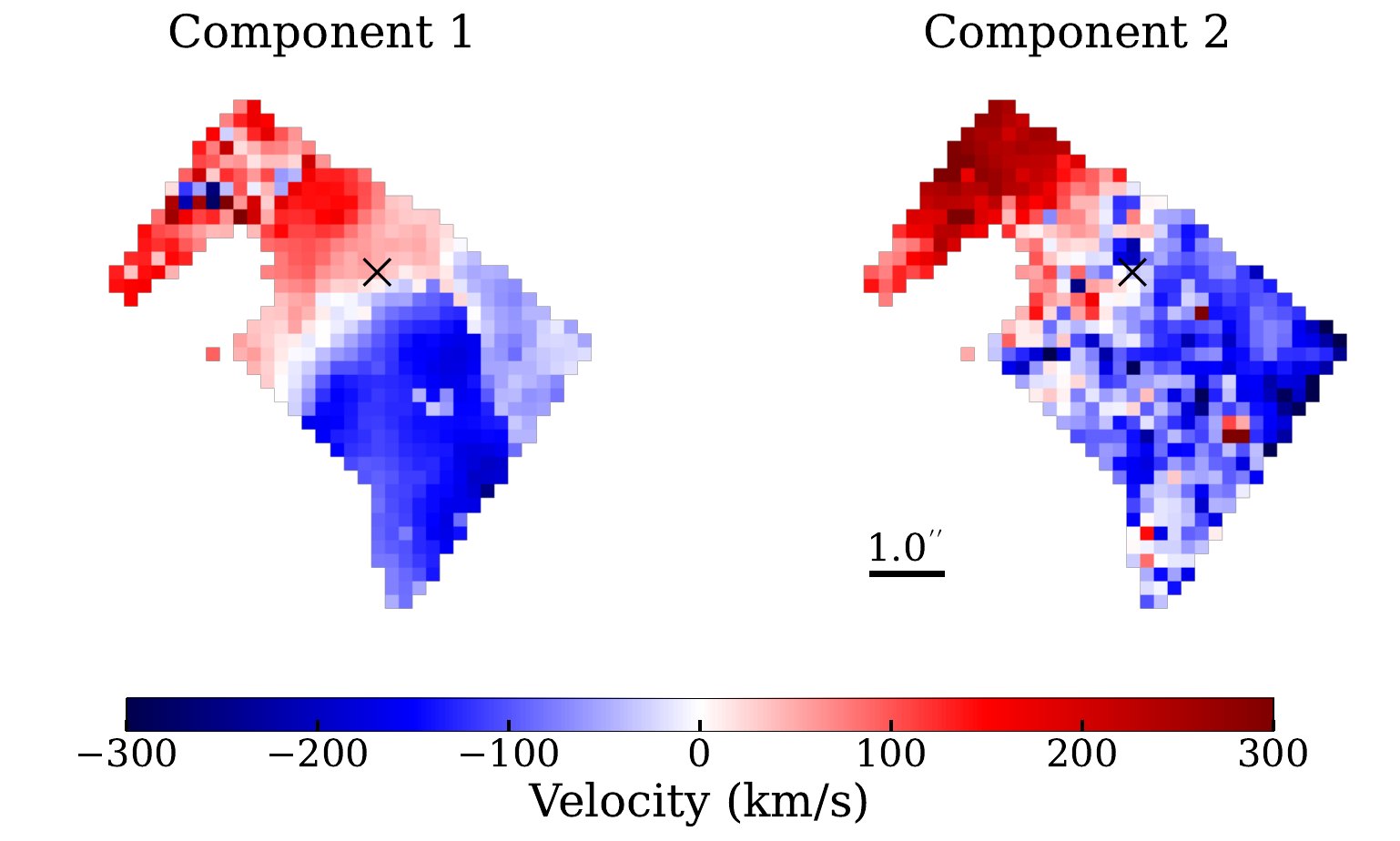}
   \caption{[S IV] velocity maps from double-Gaussian fitting. This line has an improved fit when using a double over a single-Gaussian line profile, however both velocity maps strongly trace the outflow likely due to the bicone being hollow. The same effect is seen with other strongly outflow tracing lines with high S/N, [O IV] and [Ne V].}
    \label{fig:SIV_2maps}%
\end{figure}

In contrast, Fig.~\ref{fig:SIV_spaxel} shows a clear double-peaked profile for [S IV], a feature present in most spaxels when both edges of the cone are in the line of sight, and also observed in the other high-S/N outflow tracers, [O IV] and [Ne V], and is likely due to the bicone being hollow as discussed in Section~\ref{sec:doublegauss}. These profiles are significantly better described by double-Gaussian fits than by single-Gaussian models. The two fitted components have comparable amplitudes and widths (more so than a disk component), and their corresponding velocity maps exhibit nearly identical kinematic structures, including similar major-axis PAs and velocity extrema, confirming that both components trace the outflow kinematics (see Fig.~\ref{fig:SIV_2maps}).

Finally, Fig.~\ref{fig:NeIII_spaxel} shows a representative [Ne III] spaxel spectrum (in the line of sight of both the cone and the disk) exhibiting an asymmetric profile with a pronounced wing, which is significantly better fit by a double- rather than single-Gaussian model, as indicated by a substantial reduction in the $\chi^{2}$. The two components typically have distinct parameters: a broader, lower-amplitude component tracing the outflow, and a narrower, higher-amplitude component tracing rotation in the circumnuclear disk. This behaviour is consistently observed across the IFU for [Ne III], [Ar III], and [S III], leading us to conclude that these lines robustly trace both disk and outflow kinematics.

\section{Resultant fits from disk and outflow modelling}

\begin{figure}
   \centering
   \includegraphics[width=\columnwidth]{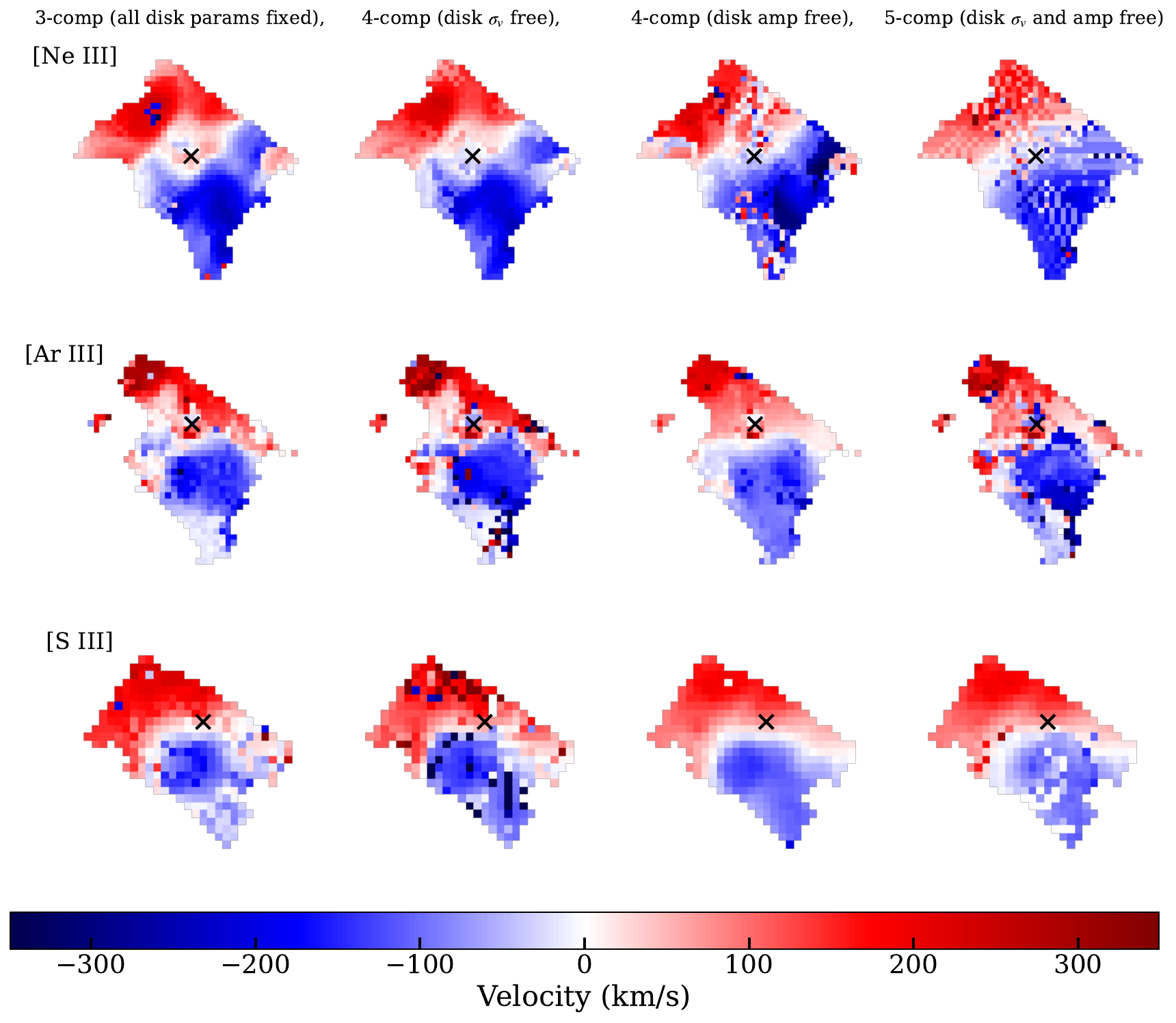}
   \caption{Resultant intermediate-IP line outflow velocity maps after incorporating the thin disk velocity model, Fig.~\ref{fig:NeII_disk_velocity_map} (top), in different ways. Left column shows the resulting single-Gaussian fit velocity map after fixing all of the disk Gaussian parameters from the thin inclined disk model. 2nd column shows the velocity map when the disk velocity dispersion is an extra free parameter. 3rd column shows the velocity map when the disk amplitude is an extra free parameter. 4th column shows the velocity map when both the disk dispersion and amplitude are both extra free parameters. Each row shows a different intermediate-IP line as given by the label on the left of each row.}
    \label{fig:cone_fits_from_disk_model}%
\end{figure}

\begin{figure}
   \centering
   \includegraphics[width=\columnwidth]{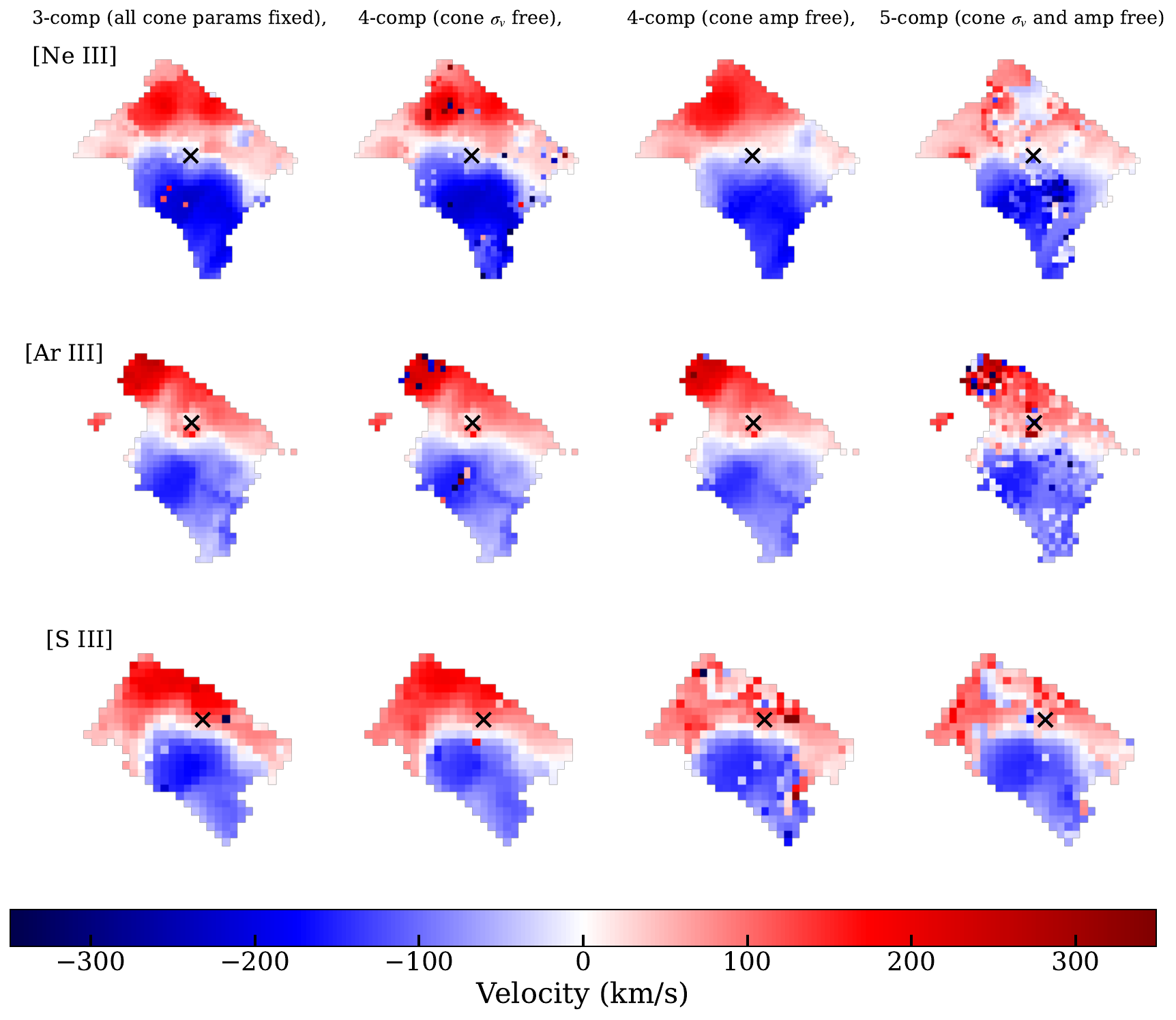}
   \caption{Resultant intermediate-IP line disk velocity maps after incorporating the simple outflow velocity model, Fig.~\ref{fig:NeII_disk_velocity_map} (bottom), in different ways. Left column shows the resulting single-Gaussian fit velocity map after fixing all of the outflow Gaussian parameters from the simple outflow model. 2nd column shows the velocity map when the outflow velocity dispersion is an extra free parameter. 3rd column shows the velocity map when the outflow amplitude is an extra free parameter. 4th column shows the velocity map when both the outflow dispersion and amplitude are extra free parameters. Each row shows a different intermediate-IP line as given by the label on the left of each row.}
    \label{fig:disk_fits_from_cone_model}%
\end{figure}

\begin{figure}
   \centering
   \includegraphics[width=\columnwidth]{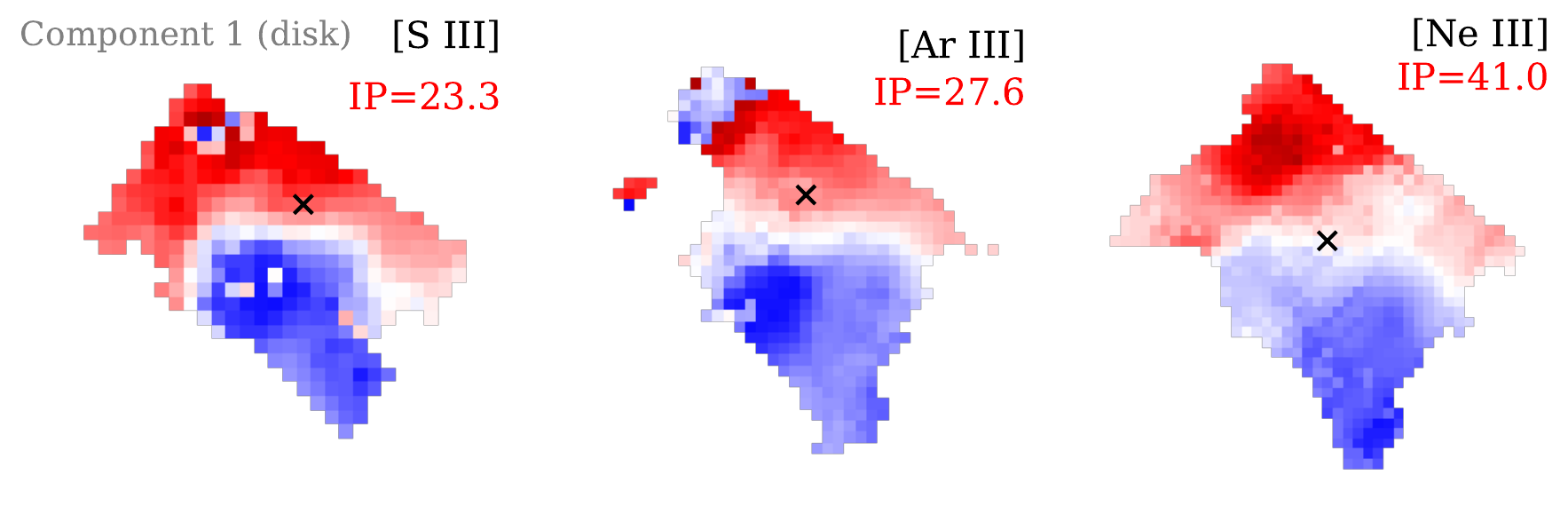}
   \includegraphics[width=\columnwidth]{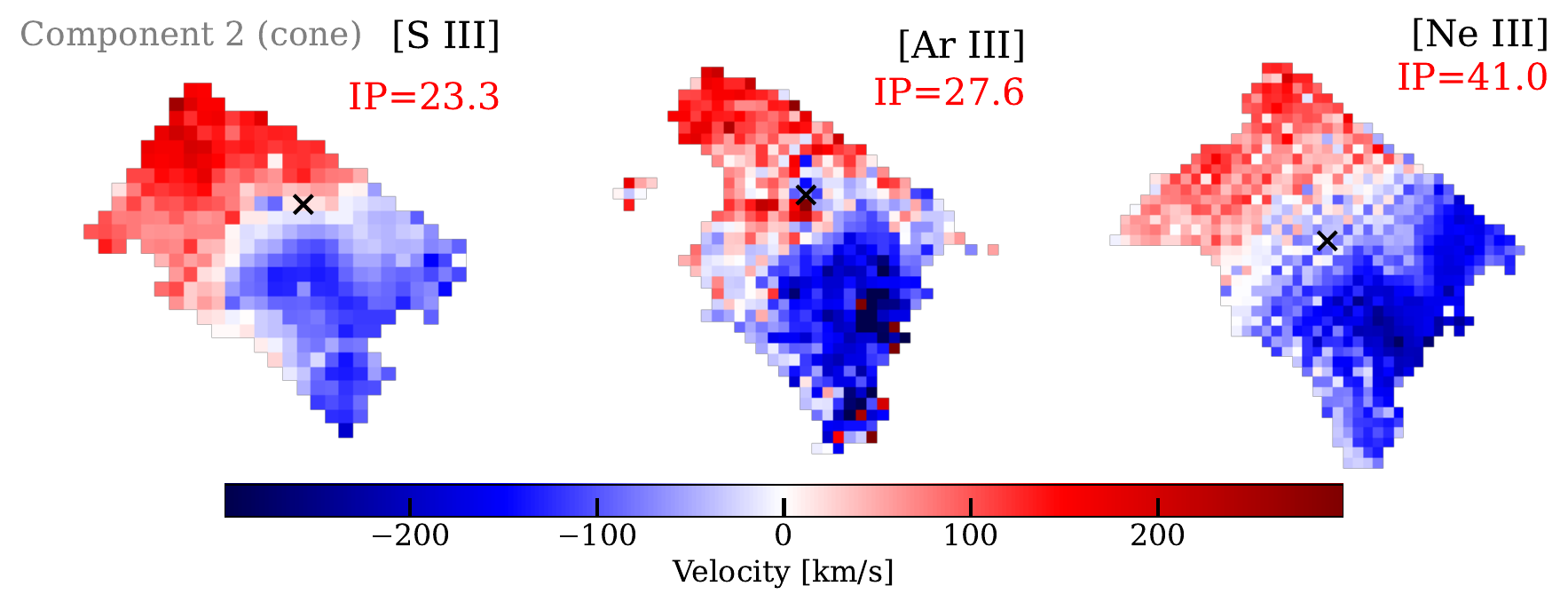}
   \caption{Velocity maps for the intermediate-IP emission line fit using double-Gaussians but with both the disk and outflow velocity dispersions fixed to the median values from Fig.~\ref{fig:double_gauss_dispersion_histograms}, showing moderate improvement on the naive double-Gaussian fits shown in Fig.~\ref{fig:double_gauss_vel_comp}. Top: disk Gaussian component, Bottom: outflow Gaussian component. The kinematics trace (disk or outflow) was set based on the criteria $\sigma_{v-\text{disk}} < \sigma_{v-\text{cone}}$ for each spaxel. Crosses mark the AGN position. North is up, east is to the left.}
    \label{fig:hybrid_double_gauss_vel_comp}%
\end{figure}

We include the resultant intermediate-IP line velocity maps after modelling a single component (disk or outflow) in different ways in Figs.~\ref{fig:cone_fits_from_disk_model} and \ref{fig:disk_fits_from_cone_model}, and from fitting a double-Gaussian with fixed velocity dispersion in Fig.~\ref{fig:hybrid_double_gauss_vel_comp}.

%%%%%%%%%%%%%%%%%%%%%%%%%%%%%%%%%%%%%%%%%%%%%%%%%%

% Don't change these lines
\bsp	% typesetting comment
\label{lastpage}
\end{document}